\documentclass[aps,prb,floatfix,twocolumn,showpacs,superscriptaddress,groupedaddress]{revtex4-2}  % for review and submission
\usepackage{graphicx}  % needed for figures
\usepackage{amsmath,amssymb}
\usepackage{flushend}
\usepackage{array}
\usepackage{float}
\usepackage{natbib}
\usepackage{booktabs}
\usepackage[autostyle, english = american]{csquotes}
\MakeOuterQuote{"}
\usepackage{xcolor}
\usepackage{hyperref}
\usepackage{color}
\begin{document}

\title{Modulation of Polarization and Metallicity in Janus Sliding Ferroelectrics}

\author{Akshay Mahajan} \email[]{akshaymahaja@iisc.ac.in} \affiliation{Solid State and Structural Chemistry Unit,
Indian Institute of Science, Bangalore 560012, India.}
\author{Awadhesh Narayan} \email[]{awadhesh@iisc.ac.in} \affiliation{Solid State and Structural Chemistry Unit,
Indian Institute of Science, Bangalore 560012, India.}

\vskip 0.25cm

\date{\today}

\begin{abstract}

Sliding ferroelectricity is emerging as a distinct and promising mechanism for realizing ferroelectricity in low-dimensional systems, offering new design principles beyond the conventional ferroelectric mechanism. Further, the coexistence of the out-of-plane polarization with in-plane conductivity induced by electrostatic charge doping makes these systems strong candidates for realizing ferroelectric metals. Using density functional theory calculations, we analyze the transition metal dichalcogenides (TMDs) based Janus sliding ferroelectric bilayers XMY (M = Mo, W; X, Y = S, Se, Te; X $\neq$ Y). In addition to exhibiting switchable interlayer polarization, Janus sliding ferroelectrics possess an intrinsic electric field within each monolayer, arising from the electronegativity difference between the chalcogen atoms. We discover that the intrinsic electric field of the monolayers can be used to modulate the interlayer ferroelectric polarization and the electronic band structure. We identify the decrease in the interlayer distance due to a particular stacking of the Janus bilayers as a major contributor to increasing polarization and reducing the bandgap. The direction of the intrinsic electric field within the Janus monolayers plays a significant role in the modulation of layer-wise contribution in the valence and conduction bands, which influences the polarization reduction due to extrinsic charge dopants. Extending this concept to Janus trilayers, we observe further enhancement in polarization and additional bandgap reduction compared to their bilayer counterparts. These results highlight the tunability of TMD-based Janus sliding ferroelectrics and suggest a pathway for designing low bandgap ferroelectrics and potential ferroelectric metals.          
\end{abstract}

%\pacs{}
\maketitle

%\end{titlepage}

\section*{Introduction}
\label{intro}

In recent years, the focus of the ferroelectric research community has shifted toward two-dimensional (2D) van der Waals (vdW) ferroelectrics for nanoscale device applications, owing to their atomic-scale thickness, clean surfaces, ease of integration, and high dielectric constants~\cite{Xiao2019,MenghaoWu2021}. Several promising candidates have been identified through first-principles simulations~\cite{Kruse2023,Xiao2019}. Experimental demonstrations of ferroelectricity have included both in-plane polarization~\cite{Chang2016,Chang2020,Higashitarumizu2020,Ghosh2019,Gou2023} and, in some materials, the more desirable out-of-plane polarization~\cite{Liu2016,Belianinov2015,Cui2018,Zhou2017,Yuan2019}. The coexistence of metallicity and switchable out-of-plane polarization remains exceptionally rare and has been observed primarily in semi-metallic 1T$'$-WTe$_2$ bilayer and multilayer systems~\cite{Fei2018}. To expand the functional landscape of low-dimensional ferroelectrics, new material designs are needed that offer out-of-plane polarization along with other useful features -- such as a low bandgap for solar energy harvesting~\cite{Zhang2018,Wu2022}, or metallic/semi-metallic behavior to explore quantum phenomena~\cite{Bhowal2023,Zhou2020,Young2023}.  

In 2017, Li \textit{et al.} proposed an alternative mechanism for realizing vertical ferroelectric polarization in bilayers of certain 2D vdW materials by altering their stacking order through interlayer sliding~\cite{Li2017}. This mechanism, known as \textit{sliding ferroelectricity}~\cite{Menghao2021}, has since been experimentally confirmed in bilayers of hexagonal boron nitride (\textit{h}-BN)~\cite{Yasuda2021,Stern2021} and transition metal dichalcogenides (TMDs)~\cite{Wang2022,Deb2022}. Owing to its interfacial nature, this stacking-engineered ferroelectric polarization can coexist with electrostatically doped in-plane conductivity~\cite{Deb2022,Wang2024}, opening avenues for engineering ferroelectric metals. Interestingly, a similar mechanism accounts for the coexistence of semi-metallicity and out-of-plane polarization in 1T$'$-WTe$_2$~\cite{Yang2018}. Sliding ferroelectricity has also been demonstrated in MoS$_2$ trilayers~\cite{Deb2022,Meng2022,Cao2024}, where the presence of multiple polarization states offers prospects for designing multistate ferroelectric devices~\cite{Meng2022}. Due to the interfacial nature of sliding ferroelectrics, the polarization from both interfaces in trilayer systems contributes cumulatively, enhancing the overall polarization~\cite{Deb2022,Cao2024}. More recently, first-principles calculations have predicted sliding ferroelectricity in Janus TMD bilayers~\cite{Lin2024,Zhang2024}, further expanding the design landscape for 2D vdW ferroelectric materials by leveraging the unique properties of Janus structures~\cite{Zhang2020}.

In this work, we investigate the rhombohedral-stacked bilayers of TMDs (MX$_2$; M = Mo, W; X = S, Se, Te) along with their Janus counterparts (XMY; M = Mo, W; X, Y = S, Se, Te; X $\neq$ Y) which exhibit switchable out-of-plane polarization. Our first-principles analysis reveals that the Janus TMD-based sliding ferroelectrics exhibit higher polarization and reduced bandgap compared to their parent TMD bilayers when the intrinsic electric field corresponding to the Janus monolayer points inwards, i.e., towards the interface of the bilayer, instead of pointing outwards away from the interface. We identify the modulation of the interlayer distance due to the intrinsic electric field of the Janus monolayers as the key mechanism behind the polarization and electronic bandgap tuning in Janus sliding ferroelectrics. Our band structure analysis reveals the dependence of the layer-wise contribution of the conduction and valence bands on the direction of the electric field within the Janus monolayers, which we find has a significant effect on the charge-doping-induced depolarization in these systems. The bandgap and depolarization modulation signify the importance of Janus monolayers in designing low-dimensional ferroelectric metals. We also analyze the effect of modulation of interlayer distance on the polarization and band structures of TMD-based sliding ferroelectric bilayers. Finally, we extended the Janus sliding ferroelectrics design concept to trilayer systems by substituting the outer chalcogen atoms of the trilayer MoS$_2$ with Se or Te. The Janus trilayer systems also exhibit polarization enhancement along with bandgap reduction. Overall, our study provides comprehensive insights into tuning the ferroelectric and electronic properties of TMD-based sliding ferroelectrics and highlights Janus monolayers as promising building blocks for designing next-generation low-bandgap ferroelectrics and ferroelectric metals.    

\section*{Computational Methods}
\label{compdetails}

Our first-principles calculations were carried out within the density functional theory (DFT) formalism, as implemented in the {\sc quantum espresso} code~\cite{Giannozzi2017,Giannozzi2009}. We used ultrasoft pseudopotentials~\cite{Vanderbilt1990} with a kinetic energy cut-off of 90 Ry for the wavefunctions and 720 Ry for charge density, with a $k$-mesh of $14\times14\times1$. We employed the Perdew-Burke-Ernzerhof (PBE) formalism to describe the exchange-correlation functional~\cite{Perdew1996}, along with the Grimme-D3 dispersion correction with BJ damping~\cite{Grimme2010}. We used a vacuum of around 30\AA to avoid interaction between adjacent images due to the periodic boundary conditions. The electronic self-consistent cycle used the convergence criterion of $1 \times 10^{-8}$ eV. We relaxed the structures until the residual forces on the atoms were less than 0.001 eV/\AA. We calculated the electric polarization values using the Berry phase method~\cite{King1993,Resta1994}, unless specified otherwise. The electrostatic potential profiles were obtained using the ionic and Hartree contributions from the charge density~\cite{Wang2022,Ferreira2021}. We also used a dipole moment correction along the out-of-plane direction for the electrostatic potential profiles.

We implemented charge doping calculations using the fractional nuclear charge pseudo atom approach~\cite{Sinai2013}. Following a similar approach as Ref.~\cite{Deb2022}, we carried out the doping calculations by charging the metal nuclei. The charging of the metal nuclei was done by adding an equal amount of fractional charge to the metal nuclei of both layers for all the bilayer systems. The maximum doping limit for electron and hole dopants we used in this study is $\sim$10$^{13}$ cm$^{-2}$, which is the highest reported experimentally accessible charge density in such systems~\cite{Deb2022}. At each doping stage, we relaxed the bilayers keeping the lattice constants fixed. We used Fermi-Dirac smearing for the band structure and density of states (DOS) calculations and for the self-consistent cycle during the charge doping to enhance the convergence. We employed spin-orbit interactions for band structure, DOS, and charge doping calculations, as well as other cases as explicitly mentioned in the results section. We carried out phonon calculations within the finite displacement approach using a $2\times2\times1$ supercell, as implemented using a combination of {\sc quantum espresso} and {\sc phonopy} codes~\cite{phonopy-phono3py-JPCM,phonopy-phono3py-JPSJ}.

\section*{Results and Discussion}
\label{results}

\subsection*{Bilayer Janus Sliding Ferroelectrics}
\label{bi}
\subsubsection*{\textbf{Structure and Stability}}
\label{strucbi}

%%%% Figure 1 %%%%
\begin{figure}[t]
\includegraphics[width=\linewidth]{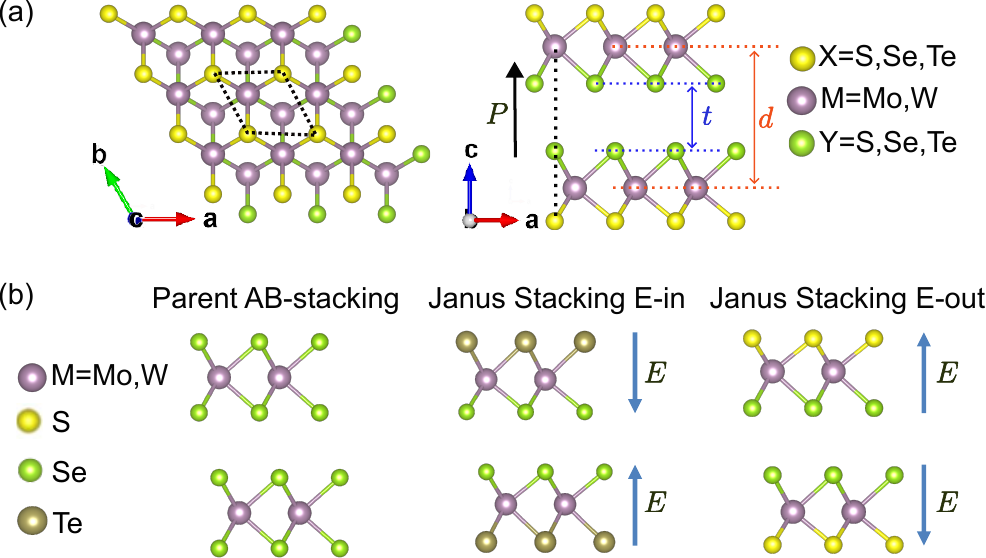}
\caption{\textbf{Crystal structure of the bilayer Janus sliding ferroelectrics.}(a) Top and side view of the AB-stacked XMY bilayers. The interlayer distance, $d$, and the interface separation, $t$, are represented by the red and blue colors in the side view, respectively. The black arrow depicts the direction of polarization. The region enclosed by dashed black lines represents the unit cell in the top view. The dashed black line in the side view allows for better visualization of the AB stacking. Here, $c$ is along the stacking direction ($z$-axis). The $a$ and $b$ are along the in-plane lattice vectors. (b) Different stacking of the intrinsic electric field, $E$, of the Janus monolayers, with AB-stacked parent bilayer shown for reference. Note that the direction of $E$ is from a lower electronegativity atom to a higher electronegativity atom.}
\label{f1}
\end{figure}
%%%% Figure 1 %%%%

The structure of the rhombohedral stacked bilayers of TMDs is shown in Figure~\ref{f1}(a). The structures are formed by stacking the 1H-XMY monolayers in parallel in the AB-stacking order, where X and Y denote the chalcogen atoms (S, Se, Te) and M denotes the metal atom (Mo, W). Here, X denotes the chalcogen atoms on the surface, while Y denotes the chalcogen atoms at the interface. In the AB stacking order, every M atom of the top layer lies over the bottom layer's chalcogen atom (Y), resulting in an out-of-plane polarization towards the upper layer. In this study, the interlayer distance, denoted by $d$, is calculated as the distance between the layers of M atoms of the top and the bottom layers. In addition, the interface distance, denoted by $t$, is calculated as the distance between the nearest neighboring atomic layers, the layer of Y atoms, of the top and bottom layers.

For the Janus systems, the layers are stacked such that the intrinsic electric field, $E$, of the Janus monolayers, will either point inward, i.e., towards the interface or will point outwards, i.e., towards the vacuum, as shown in Figure~\ref{f1}(b), resulting in E-in and E-out stacking for Janus bilayers. Other stacking orders, where the field $E$ points in the same direction in both the monolayers, will not result in sliding ferroelectricity in these Janus systems, as shown in a previous work~\cite{Lin2024}, and hence are excluded in this study.

To evaluate the energetic stability of the Janus sliding ferroelectrics, we calculated the formation energy, $E_F$, for all the structures, as shown in Figure S1~\cite{SuppMat}. The formation energy is calculated as the difference between the energy of the material system and the sum of the energies of the constituent atoms in their bulk phases. For the bilayers, this is given by,
\begin{equation}
E_F = \frac{E_{total} - E_M \times N_M - E_X \times N_X -E_Y \times N_Y}{N}. 
\label{eq1}
\end{equation}
Here, $E_{total}$ is the total energy of the XMY sliding ferroelectric bilayer; $E_M$, $E_X$, and $E_Y$ are the energies of M, X, and Y atoms in their respective bulk phases; $N_M$, $N_X$, and $N_Y$ are the number of M, X and Y atoms in the unit cell; and $N$ is the total number of atoms present in the unit cell.

All the Janus bilayers show negative formation energies, confirming their energetic stability. We also compared the formation energies of the Janus bilayers to those of the experimentally synthesized systems (1T$'$-WTe$_2$ bilayer, Janus MoSSe monolayer, and AB-stacked WSe$_2$ bilayer) to further confirm their stability, as shown in Figure S1~\cite{SuppMat}. The crystal structure parameters and the formation energies for all the structures are provided in Table S1~\cite{SuppMat}.

\subsubsection*{\textbf{Polarization Modulation}}
\label{polbi}

%%%% Figure 2 %%%%
\begin{figure*}[t]
\begin{center}
\includegraphics[width=0.75\linewidth]{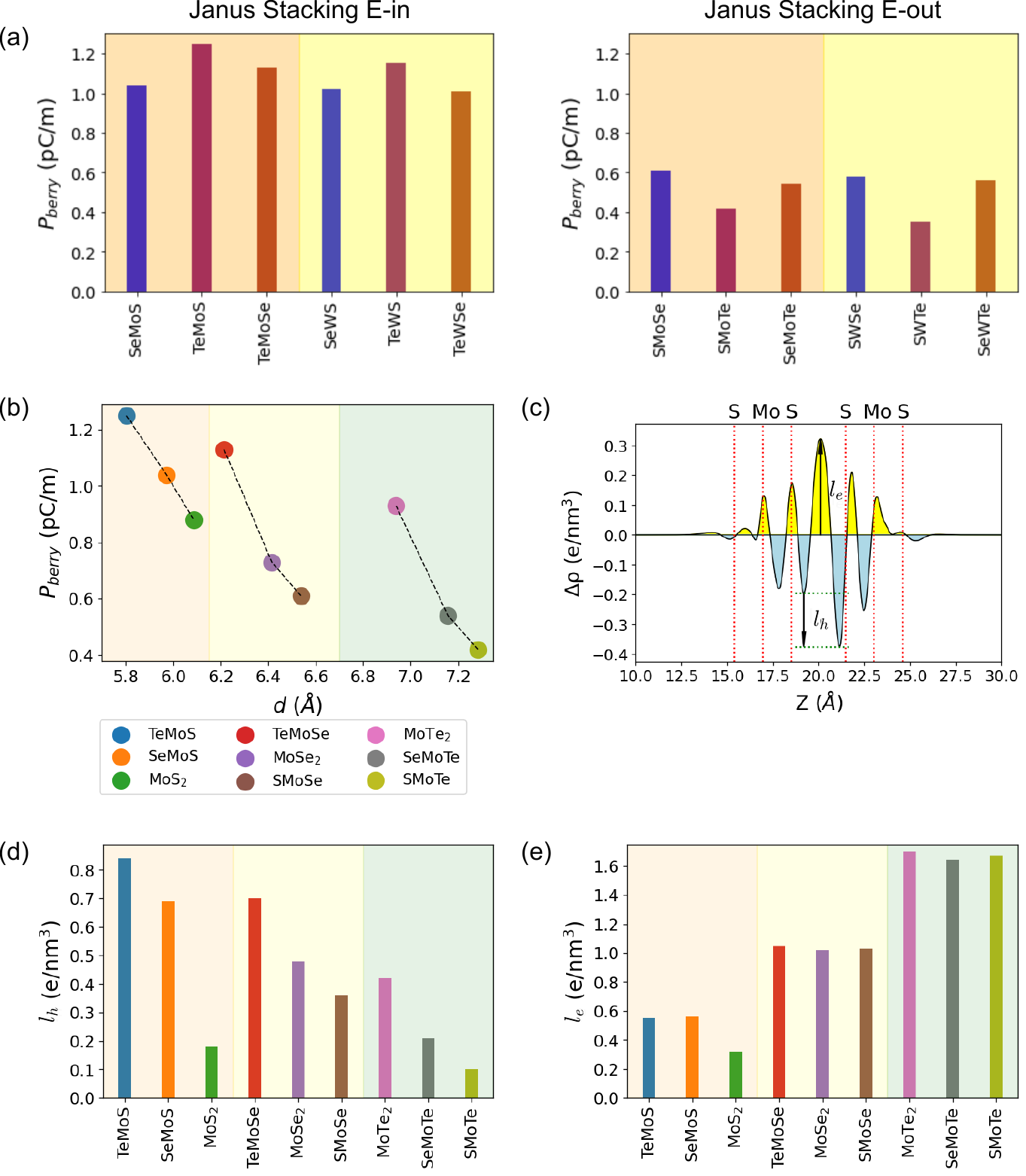}
\caption{\textbf{Polarization and interlayer charge density modulation in bilayer Janus sliding ferroelectrics.}(a) Comparison of $P_{berry}$ (polarization via Berry phase method) for the Janus E-in (left) and E-out (right) bilayers. (b) $P_{berry}$ versus interlayer distance $d$ for XMoY bilayers. (c) Charge density difference profile for MoS$_2$ bilayer. The yellow and blue colors represent the electron and hole accumulation, respectively. The dashed red lines represent the vertical location of the atoms. Black arrows denote the parameters $l_h$ (difference of hole accumulation peaks) and $l_e$ (electron accumulation peak) of the charge density profile. The variation of (d) $l_h$ and (e)$l_e$ parameters of the charge density profiles for the XMoY bilayers.}  
\label{f2}
\end{center}
\end{figure*}

Next, we analyze and compare the polarization values of the Janus sliding ferroelectric bilayers, as shown in Figure~\ref{f2}(a). The polarization values are higher for E-in stacking and lower for E-out stacking of the Janus bilayers, suggesting polarization modulation due to the $E$ fields within the Janus monolayers. In general, the E-in Janus counterparts of TMD bilayers exhibit enhanced polarization, whereas the E-out stacked Janus bilayers show reduced polarization compared to their parent TMD bilayers, as summarized in Table S2~\cite{SuppMat}. A similar polarization modulation was also observed in a previous work~\cite{Zhang2024}. Along with polarization via the Berry phase method~\cite{King1993,Resta1994}, $P_{berry}$, we also calculated the interlayer potential difference, $\Delta\phi$, and polarization values, $P_{CD}$, from the electrostatic potential and the charge density profiles, respectively. Both $\Delta\phi$ and $P_{CD}$ exhibit modulation similar to $P_{berry}$ values for Janus bilayers, as shown in Table S2~\cite{SuppMat}. 

The electrostatic potential and charge density profiles are obtained by calculating the differences between the plane-averaged potential and charge density, respectively, along the $z$-direction for the bilayer and the corresponding isolated monolayers. The difference between the electrostatic potential values obtained far (around 10\AA) above the top and below the bottom surfaces of the bilayer provides $\Delta\phi$, similar to Ref.~\cite{Deb2022}. The methodology for calculating $P_{CD}$ value is provided in the supplementary material~\cite{SuppMat}. Using the electrostatic potential and charge density profiles, we can also show the polarization direction switching with a sliding motion, as shown in Figure S2~\cite{SuppMat}. Note that the calculations of $\Delta\phi$ and $P_{CD}$ are computationally less demanding than that of $P_{berry}$ directly.

To understand the mechanism behind the modulation of polarization due to the E fields, we analyze the XMoY sliding ferroelectrics in detail. First, we compare the XMoY bilayers' polarization values concerning their interlayer distance, $d$, as shown in Figure~\ref{f2}(b). We observed that as $d$ decreases, the polarization value increases, especially when comparing the systems with the same Y atoms. This can be attributed to the enhanced interlayer vdW interactions resulting from the reduced interlayer distance ($d$), leading to a higher polarization. 

Interestingly, the E-in stacking results in a lower $d$ while the E-out stacking results in a higher $d$ than the parent bilayers. This suggests the interlayer distance modulation via the intrinsic electric field and, as a result, the modulation of polarization. Another interesting aspect is that the $d$ values mainly depend on the Y atom and change only slightly with the change in the X atom. As the Y atom size increases or the Y atom electronegativity decreases, $d$ increases, thereby decreasing the polarization.

To gain deeper insight into the modulation of interlayer distance and the resulting polarization, we further analyze the charge density profiles of the XMoY bilayers. The charge density profile of the MoS$_2$ bilayer is shown in Figure~\ref{f2}(c), where two important features in the interlayer region of the plot are the difference between the hole accumulation peaks, $l_{h}$, and the height of the electron accumulation peak, $l_{e}$. The $l_h$ parameter can be used to quantify the charge redistribution between the layers because of the interlayer interaction, which is responsible for the magnitude of the out-of-plane polarization~\cite{Zhang2024}. The $l_e$ parameter, as we have observed, correlates with the interlayer distance. The charge density profiles for all the XMoY bilayers are provided in Figure S3~\cite{SuppMat} for comparison.

The $l_h$ parameter exhibits a trend similar to that of polarization for systems containing the same Y atom, as shown in Figure~\ref{f2}(d). The $l_e$ parameter increases notably with the atomic size of Y, as illustrated in Figure~\ref{f2}(e), which accounts for the larger interlayer distances observed in the XMoTe systems in Figure~\ref{f2}(b).

To assess the role of relativistic effects on out-of-plane polarization, we compared the charge density and electrostatic potential profiles of XMoY bilayers with and without spin–orbit coupling (SOC), as shown in Figure S3 and Figure S4~\cite{SuppMat}, respectively. The profiles exhibit no significant differences, indicating minimal impact of SOC on ferroelectric polarization. A quantitative comparison of $\Delta\phi$ and $P_{CD}$ values, presented in Figure S5~\cite{SuppMat}, further confirms that SOC introduces only negligible changes, and thus does not substantially affect the switchable out-of-plane polarization in these systems.

%%%% Figure 3 %%%%
\begin{figure*}[t]
\begin{center}
\includegraphics[width=0.75\linewidth]{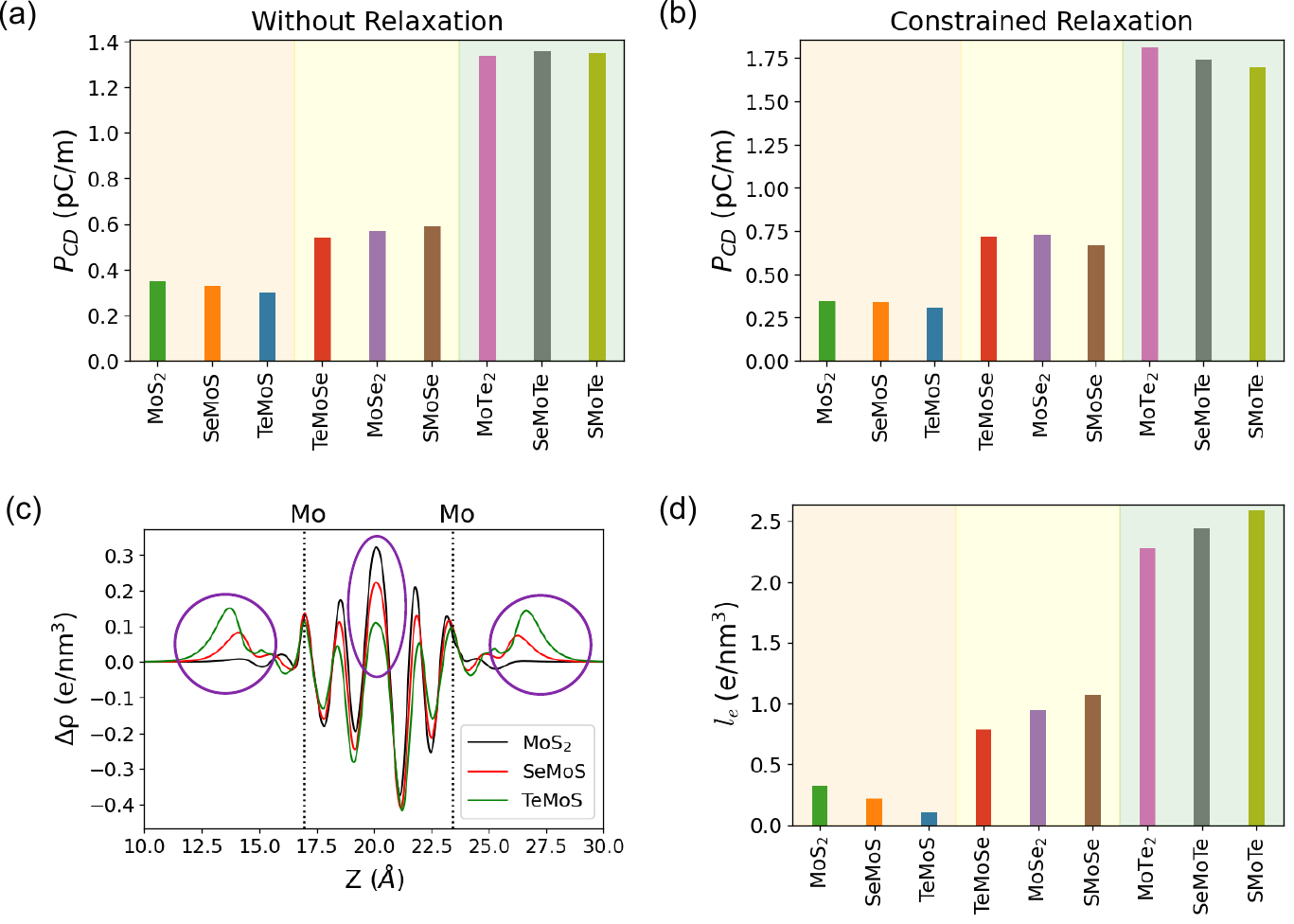}
\caption{\textbf{Polarization and interlayer charge density variation for fixed interlayer distance.} Variation of polarization calculated from charge density, $P_{CD}$, (a) without relaxation, and (b) constrained relaxation of other structural parameters. (c) Charge density profiles for XMoS bilayers. The purple circle highlights the variation in electron accumulation in the regions. (d) Comparison of the parameter $l_e$ of the charge density profiles of XMoY bilayers.}  
\label{f3}
\end{center}
\end{figure*}
%%%% Figure 3 %%%%

To further validate the proposed mechanism of polarization modulation within Janus sliding ferroelectrics, we analyze all the XMoY bilayers with their interlayer distance ($d$) fixed to the average interlayer distance ($d_{avg}$) of the nine systems, i.e., $d = d_{avg} = 6.49$\AA. For this computational experiment, we first fixed $d$ of the MoS$_2$ (SMoS) bilayer to $d_{avg}$ and relaxed all the other structural parameters of the system. Then, we created all the other XMoY structures from the relaxed MoS$_2$  bilayer with $d = d_{avg}$, via chemical substitution. The obtained bilayers have the same structural parameters, and the $P_{CD}$ values for them are provided in Figure~\ref{f3}(a). Next, we relaxed all structural parameters except $d$, which led to a slight decrease in the interface distance ($t$) for bilayers with Y = Se or Te, as shown in Table S3. This reduction in $t$ results in a modest increase in the $P_{CD}$ values for these bilayers, as shown in Figure~\ref{f3}(b).  

Remarkably, there is no effective polarization modulation for the case of $d=d_{avg}$, with or without constrained relaxation, due to the stacking of $E$ fields in the Janus sliding ferroelectrics. Instead, we find that the size of the Y atom predominantly affects the polarization, resulting in higher polarization for larger Y atoms, irrespective of the X atom. These findings affirm the proposed mechanism of polarization modulation within Janus sliding ferroelectrics via interlayer distance tuning. In addition, considering the same $d$ values, bilayers with larger Y atoms exhibit higher polarization values.

The analysis of the charge density profiles for the structures with $d=d_{avg}$ explains how the modulation of the interlayer charge distribution via the E fields in the Janus sliding ferroelectrics results in the modulation of $d$. As seen in Figure~\ref{f3}(c), the E-in stacking reduces the electron accumulation between the bilayers by redistributing it towards the outer surface. As a result, the $l_e$ parameter decreases, as shown in Figure~\ref{f3}(d), which is expected to result in a reduction of the interlayer distance. Similarly, the E-out stacking results in an increased electron accumulation between the layers as shown in Figure S6\cite{SuppMat} and depicted from the increase in $l_e$ in Figure \ref{f3}(d), resulting in a larger interlayer distance on complete relaxation of the bilayers. Thus, the E fields within the Janus monolayers tune the electron accumulation within the interlayer region (or the $l_e$ parameter of the charge density profile), which modulates the interlayer distance of Janus sliding ferroelectrics.     

Additionally, the higher values for $l_h$ in Figure S6~\cite{SuppMat} for the XMoTe bilayers explain their higher polarization values for the case of $d=d_{avg}$. Thus, for the same $d$ values, the larger Y atoms result in stronger interlayer interactions, which explains the higher polarization values for XMoTe bilayers in Figure~\ref{f3}(a)-(b).      

Finally, to validate our analysis, we repeated the polarization calculations -- using $P_{CD}$ and $\Delta\phi$ -- for the structures analyzed in Figure~\ref{f3}(b), varying the interlayer distance $d$ around $d_{avg}$, as presented in Figure S6~\cite{SuppMat}. No effective polarization modulation was observed via the $E$ fields within Janus sliding ferroelectrics for fixed $d$.

\subsubsection*{\textbf{Band Structure Modulation}}
\label{bandbi}

%%%% Figure 4 %%%%
\begin{figure*}[t]
\begin{center}
\includegraphics[width=0.8\linewidth]{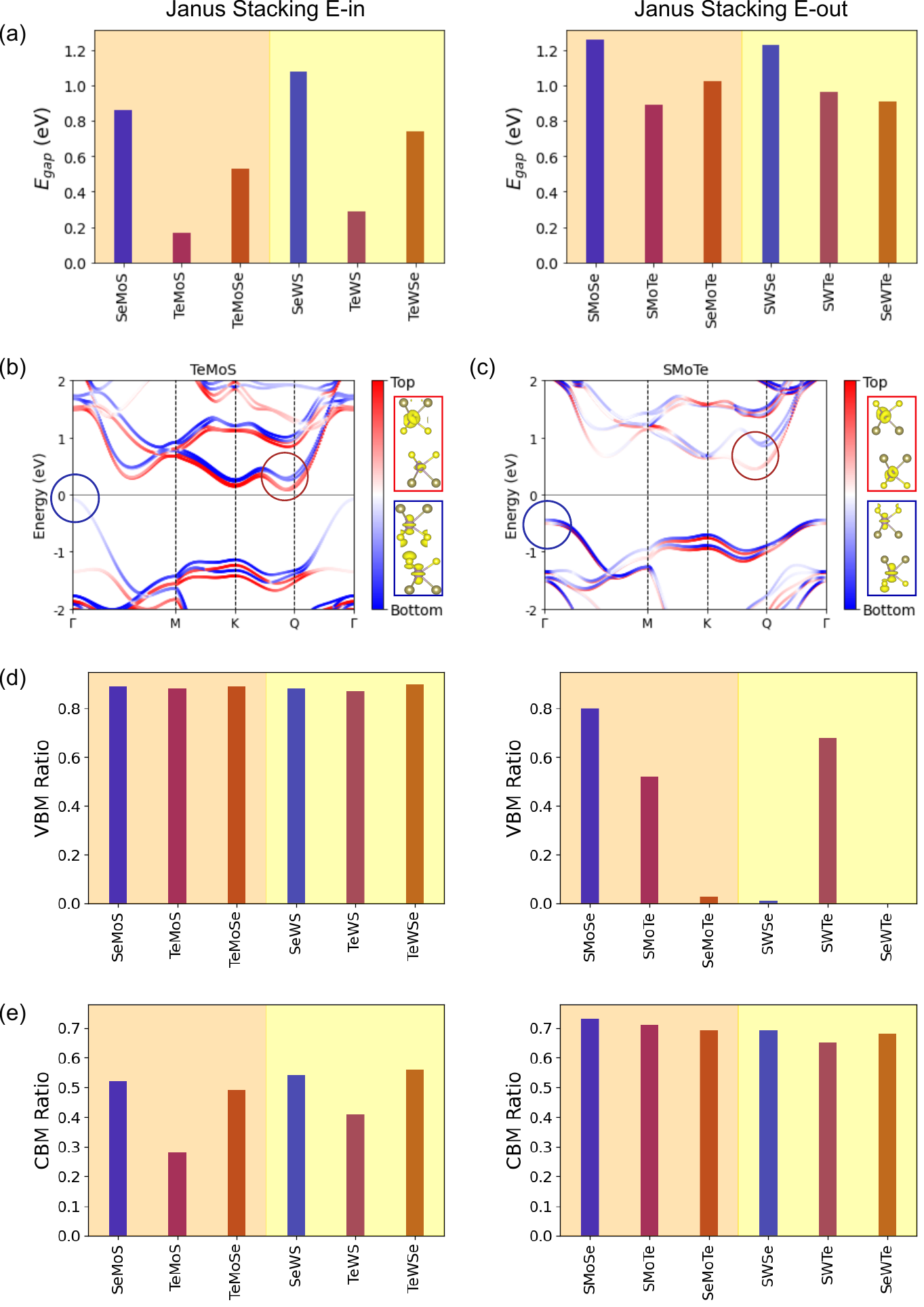}
\caption{\textbf{Band structure modulation in bilayer Janus sliding ferroelectrics.}(a) Comparison of electronic bandgap, $E_{gap}$, for the Janus E-in (left) and E-out (right) bilayers. Layer-contribution-projected band structure of (b) TeMoS and (c) SMoTe bilayers. The blue and red circles represent the region around the valence band maximum (VBM) and the conduction band minimum (CBM), respectively. The blue and red colors represent the contribution of the bottom and top layers to the particular band, respectively. Here, Q is the midpoint between K and $\Gamma$. The zero of the energy scale is at the Fermi level. The illustrations on the top right (red color box) and on the bottom right (blue color box) are the partial charge density at the CBM and VBM, respectively. (d) VBM ratio and (e) CBM ratio for the Janus E-in (left) and E-out (right) bilayers.}  
\label{f4}
\end{center}
\end{figure*}
%%%% Figure 4 %%%%

Following the polarization analysis, we examine the electronic band structures of the two Janus stacking configurations in Figure~\ref{f4}. The E-in stacking, shown in Figure~\ref{f4}(a), exhibits a lower bandgap than E-out stacking. This demonstrates that the bandgap can be tuned by altering the intrinsic electric field ($E$) configuration in Janus sliding ferroelectrics. Compared to the parent TMDs, the E-in stacked Janus structures exhibit reduced bandgaps, as shown in Table S2~\cite{SuppMat}. This highlights a novel approach to achieving low-bandgap ferroelectrics through E-in stacking of Janus counterparts of TMD-based sliding ferroelectrics.

Next, we analyze the layer-resolved contributions to the valence and conduction bands to understand how the stacking sequence in Janus bilayers influences band structure modulation. We begin by comparing the layer-contribution-projected band structures of TeMoS (E-in stacking) and SMoTe (E-out stacking) bilayers, shown in Figure~\ref{f4}(b)-(c). At the valence band maximum (VBM), the TeMoS bilayer exhibits a more balanced contribution from the top and bottom layers than SMoTe. In contrast, at the conduction band minimum (CBM), the SMoTe bilayer shows a more equal layer contribution than TeMoS. Similar trends are observed across other XMoY bilayers, as illustrated in Figure S7~\cite{SuppMat}. These results indicate that the $E$ fields within Janus monolayers modulates the layer-wise distribution of the VBM and CBM in Janus sliding ferroelectrics.

To quantify the asymmetrical contribution of the top and bottom layers to VBM and CBM, we define the VBM and CBM ratio for the AB-stacked bilayers as follows,

\begin{equation}
\text{VBM ratio} = \frac{C_{top}^{\text{VBM}}}{C_{bottom}^{\text{VBM}}},
\label{eqVBMratio}
\end{equation}
\begin{equation}
\text{CBM ratio} = \frac{C_{bottom}^{\text{CBM}}}{C_{top}^{\text{CBM}}}.
\label{eqCBMratio}
\end{equation}

Here $C_{top/bottom}^{\text{VBM/CBM}}$ denote the contribution of the top or bottom layer to the VBM or CBM, expressed in terms of atomic orbital projections of the wavefunctions. Thus, the VBM and CBM ratios can be used to quantitatively compare the contribution of each layer to the VBM and CBM.

To understand how these ratios are defined, we consider the direction of the interlayer electric field and its influence on the VBM and CBM contributions. The interlayer polarization points upward in the bilayers, as shown in Figure~\ref{f1}(a), resulting in a downward-pointing electric field, i.e., directed towards the bottom layer. This field increases the potential energy of electrons in the bottom layer, leading to a splitting of the VBM and CBM states. This effect is evident in the band structures of AB-stacked XMoY bilayers shown in Figure S7~\cite{SuppMat}, where the predominantly blue (bottom layer) bands appear at higher energies than the red (top layer) bands near the VBM and CBM. Consequently, the CBM and VBM show greater contributions from the top and bottom layers, respectively. A similar band alignment model near the Fermi level is proposed for bilayer sliding ferroelectrics in Ref.~\cite{Wang2024}.  

Based on the band alignment model near the Fermi level, the bottom layer contributes more to the VBM than the top layer, implying $C_{top}^{VBM}<C_{bottom}^{VBM}$. Similarly, the top layer has a greater contribution at the CBM, leading to $C_{bottom}^{CBM}<C_{top}^{CBM}$. According to Eq.~\ref{eqVBMratio} and Eq.~\ref{eqCBMratio}, the VBM and CBM ratios attain a maximum value of 1 when both layers contribute equally. Thus, higher VBM and CBM ratios indicate a more balanced contribution from the top and bottom layers.   

A comparison of the VBM and CBM ratios for the E-in and E-out stacked Janus bilayers in Figure~\ref{f4}(d)-(e) reveals that E-in stacking yields higher VBM ratios and lower CBM ratios than E-out stacking. To understand this modulation, we examine how the $E$ fields within the Janus monolayers influence band splitting and, consequently, the layer-wise contributions at the VBM and CBM. When the splitting between bands near the VBM and CBM is small, the corresponding states become more localized on a specific layer. In contrast, larger splitting results in a more delocalized distribution across layers.  

By comparing the band structures of TeMoS (E-in) and SMoTe (E-out) bilayers, shown in Figure~\ref{f4}(b)-(c), we observe that E-in stacking leads to greater splitting near the VBM and reduced splitting near the CBM. This results in more delocalized bands at the VBM, leading to higher VBM ratios. Conversely, E-out stacking shows more delocalized bands at the CBM, resulting in higher CBM ratios. Thus, the modulation of band-splitting due to E-in and E-out stacking governs the variation in VBM and CBM ratios through the $E$ fields of the Janus monolayers.

We also investigated the effect of SOC on the band structures of the AB-stacked bilayers. Although SOC typically results in a slight reduction in the bandgap, it does not significantly alter the layer-wise contributions to the VBM and CBM -- unless there is a shift in their positions. Notably, in MoTe$_2$, WSe$_2$, and SWSe bilayers, SOC  causes the VBM to shift from the $\Gamma$ point to the K point, as shown in Figure S8~\cite{SuppMat}. This shift leads to a pronounced decrease in the VBM ratio, due to the highly localized nature of the VBM at the K point, reflected by the dark-blue colored VBM. 

Since the $E$ fields within Janus monolayers modulate the interlayer distance ($d$), we examine the band structures of E-in and E-out stacked Janus XMoY bilayers with $d$ fixed to $d_{avg}$, analogous to our earlier polarization analysis. For the constrained relaxation case, we find no significant difference in the bandgaps of the E-in and E-out configurations, as shown in Table S3~\cite{SuppMat}.

We further compare the VBM and CBM ratios for the XMoY bilayers with $d=d_{avg}$ in Table S3\cite{SuppMat}. The E-in stacked Janus bilayers exhibit higher VBM ratios than their E-out counterparts. However, due to a shift in the CBM position for these Janus bilayers, a direct comparison of the CBM ratios is less straightforward. 

Overall, fixing $d$ eliminates the bandgap variation due to the $E$ fields. However, variations in the layer-wise contributions to the VBM and CBM remain observable. This indicates that, as with polarization, the bandgap modulation in Janus bilayers due to the $E$ fields primarily arises from the tuning of $d$. Nevertheless, the direction of the $E$ fields in the Janus bilayers, i.e., E-in or E-out stacking, continues to play a crucial role in determining the layer-resolved contributions to the electronic bands. Thus, E-field–based stacking of Janus sliding ferroelectrics offers an additional strategy for band structure modulation in 2D materials, complementing the more commonly employed approach of constructing versatile heterostructures~\cite{Nguyen2021,Truong2025,Nguyen2025}.

\subsubsection*{\textbf{Modulation of Doping-induced Depolarization}}
\label{dopbi}

As shown in Refs.~\cite{Deb2022,Wang2024}, the extent of polarization reduction caused by extrinsic charge doping depends on the relative contributions of each layer to the valence and conduction bands. When both layers contribute equally, the polarization reduction or the depolarization effect from electron or hole doping is minimized. Since the intrinsic electric fields ($E$) within the Janus monolayers modulate the VBM and CBM contributions in the Janus bilayers, they can also modulate the doping-induced depolarization. This, in turn, allows modulation of the coexistence of ferroelectric polarization and dopant-induced metallicity.

The effect of charge dopants on the polarization can be understood by comparing the reduction in the magnitude of polarization with charge doping relative to the undoped bilayer system. This phenomenon of polarization reduction is referred to as  'depolarization'~\cite{Deb2022}. Note that instead of polarization, we calculated the variation in the interlayer potential difference ($\Delta\phi$) with charge doping, as shown in Figure S9 and S10~\cite{SuppMat}, as polarization for metallic systems is not well-defined, following the approach used in previous studies~\cite{Deb2022,Wang2024}. Using this definition, the percentage of depolarization due to hole doping (Depol$_{hole}$) and electron doping (Depol$_{electron}$), are calculated as

\begin{equation}
\text{Depol}_{hole} = \frac{\Delta\phi_{0} - \Delta\phi_{h_{max}}}{\Delta\phi_{0}} \times 100,
\label{eqdeph}
\end{equation}
\begin{equation}
\text{Depol}_{electron} = \frac{\Delta\phi_{0} - \Delta\phi_{e_{max}}}{\Delta\phi_{0}} \times 100.
\label{eqdepe}
\end{equation}

Here, $\Delta\phi_{0}$, $\Delta\phi_{h_{max}}$, and $\Delta\phi_{e_{max}}$ are the interlayer potential differences for the AB-stacked XMY bilayers with zero doping, maximum hole doping, and maximum electron doping, respectively. 

Although one can not define polarization for metallic systems as per the Berry phase method~\cite{King1993,Resta1994}, we can get an estimate of the polarization from $P_{CD}$ values, as shown in Figure S9~\cite{SuppMat}. $P_{CD}$ shows a similar doping-induced modulation as $\Delta\phi$.  

%%%% Figure 5 %%%%
\begin{figure*}[t]
\begin{center}
\includegraphics[width=0.8\linewidth]{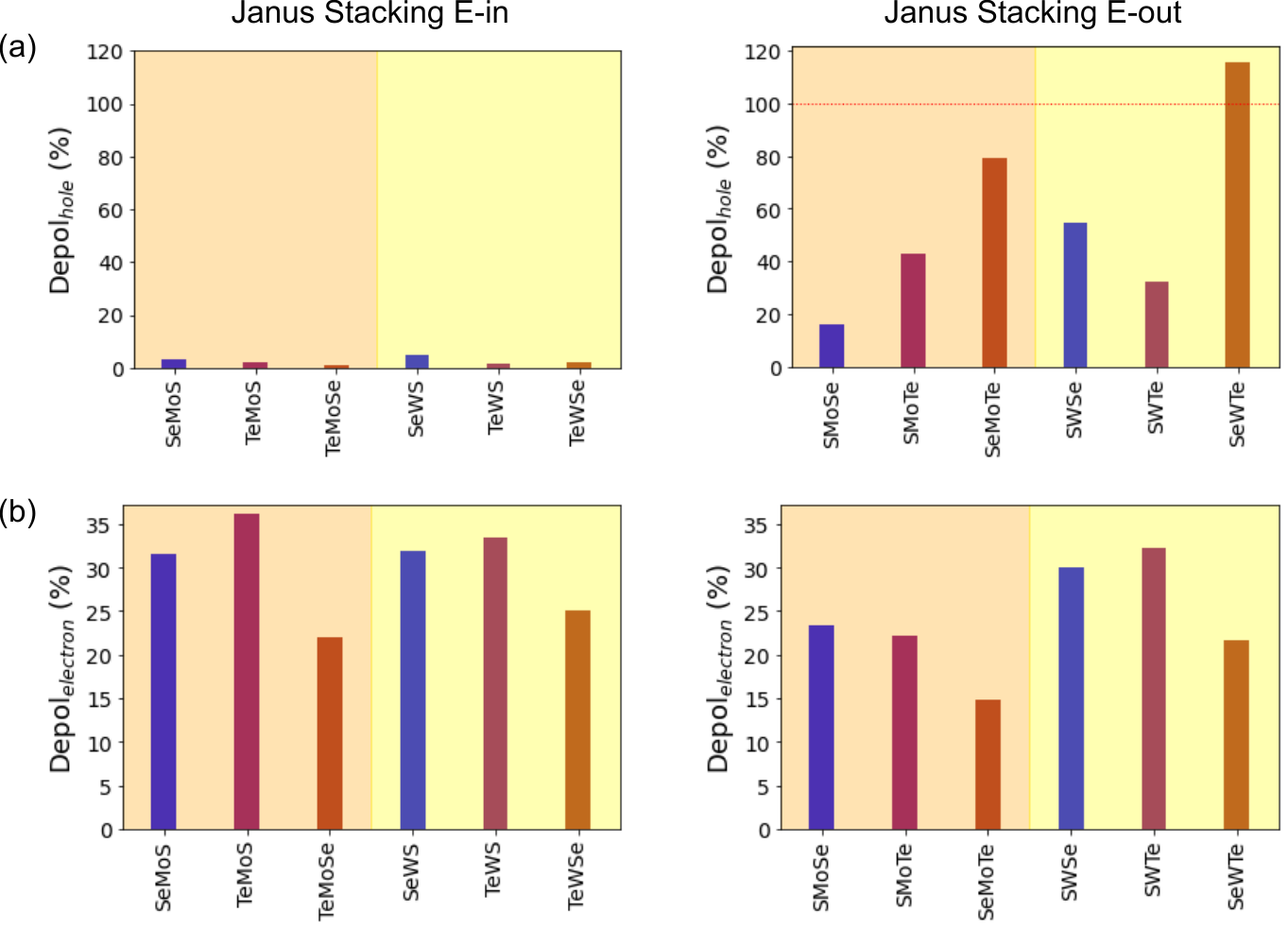}
\caption{\textbf{Modulation of doping-induced depolarization in bilayer Janus sliding ferroelectrics.} Comparison of (a) Depol$_{hole}$ and (b) Depol$_{electron}$ for the Janus E-in and E-out bilayers. The Depol$_{hole}$ crossing 100\% (red dotted line) in (a) for SeWTe suggests reversal of polarization with high doping.}  
\label{f5}
\end{center}
\end{figure*}
%%%% Figure 5 %%%%

As shown in Figure~\ref{f5}(a), under hole doping, the E-in stacked Janus bilayers exhibit significantly less depolarization compared to their E-out stacked counterparts. In contrast, under electron doping, as shown in Figure~\ref{f5}(b), the E-out stacked bilayers show lower depolarization than the E-in stacked ones, although the difference is less pronounced than in the hole doping case. 

Depolarization trends in Figure~\ref{f5} align with the VBM and CBM ratio variations shown in Figure~\ref{f4}(d)-(e) for different $E$ field stacking (E-in and E-out). Specifically, depolarization from hole and electron doping decreases as the VBM and CBM ratios increase, respectively. This correlation between doping-induced depolarization and the VBM/CBM ratios confirms that the $E$ fields in Janus monolayers modulate the charge doping-induced depolarization in Janus sliding ferroelectrics.

To clarify the conditions under which metallicity and polarization coexist in doped bilayers, we examined the variation of the Fermi energy with doping and compared the density of states (DOS) of undoped and doped TeMoS and SMoTe bilayers, as shown in Figures S11 and S12~\cite{SuppMat}. Electron (hole) doping raises (lowers) the Fermi energy, eventually shifting it across the CBM (VBM) of the undoped bilayer. For the highest doping levels, the Fermi energy falls within regions of non-zero DOS in the undoped case, and the corresponding DOS plots for doped bilayers confirm finite DOS values at the Fermi level, indicating sufficient free charge carriers for metallic behavior. Similar results are observed across other bilayers, where doping near $10^{13}$ cm$^{-2}$ consistently yields non-zero DOS at the Fermi level. Hence, bilayers doped at this level are expected to exhibit appreciable in-plane conductivity alongside switchable out-of-plane polarization, thereby fulfilling the criteria of a ferroelectric metal.

Since the doped holes and electrons populate the valence and conduction bands, respectively, their distribution in the reciprocal space is influenced by the layer-wise contribution of the VBM and CBM. This, in turn, affects the layer-wise distribution of charge carriers and the resulting charge imbalance between layers, which drives polarization, as shown in Figure S2(b)~\cite{SuppMat}. Note that a higher VBM or CBM ratio results in a more uniformly layer-wise distributed charge density at the VBM and CBM, respectively, as can be seen from the partial charge density plots at the VBM and CBM in Figure~\ref{f4}(b)-(c) and also in Figure S7~\cite{SuppMat}. This uniformity reduces the impact of doped carriers on charge redistribution and thus on polarization. 

In the E-in stacked Janus TeMoS bilayer, the CBM charge density is localized on the top layer, whereas the VBM charge density is more evenly distributed across both layers [Figure~\ref{f4}(b)]. As a result, doped electrons preferentially occupy the top layer, leading to stronger electron-induced depolarization, while doped holes are redistributed more uniformly, producing weaker hole-doping-induced depolarization. Reversing the intrinsic electric field, as in the E-out stacked Janus SMoTe bilayer, inverts the nature of charge density profiles of both the VBM and CBM. In this case, the VBM charge density localizes on the bottom layer [Figure~\ref{f4}(c)], giving rise to stronger hole-doping-induced depolarization. A more detailed analysis of the VBM and CBM layer-wise contribution and its effect on the doping-induced depolarization can be found in Ref.~\cite{Wang2024}.

We also examined the effect of SOC on the variation of $\Delta\phi$ with charge doping, as shown in Figures S9 and S10~\cite{SuppMat}. Notable differences between SOC and non-SOC results were observed only for hole doping in MoTe$_2$, WSe$_2$, and SWSe bilayers, which can be attributed to the SOC-induced shift in the VBM, as shown in Figure S8~\cite{SuppMat}.

\subsubsection*{\textbf{Polarization and Band Structure Tuning with Interlayer Distance}}
\label{interdis}

%%%% Figure 6 %%%%
\begin{figure*}[t]
\begin{center}
\includegraphics[width=0.8\linewidth]{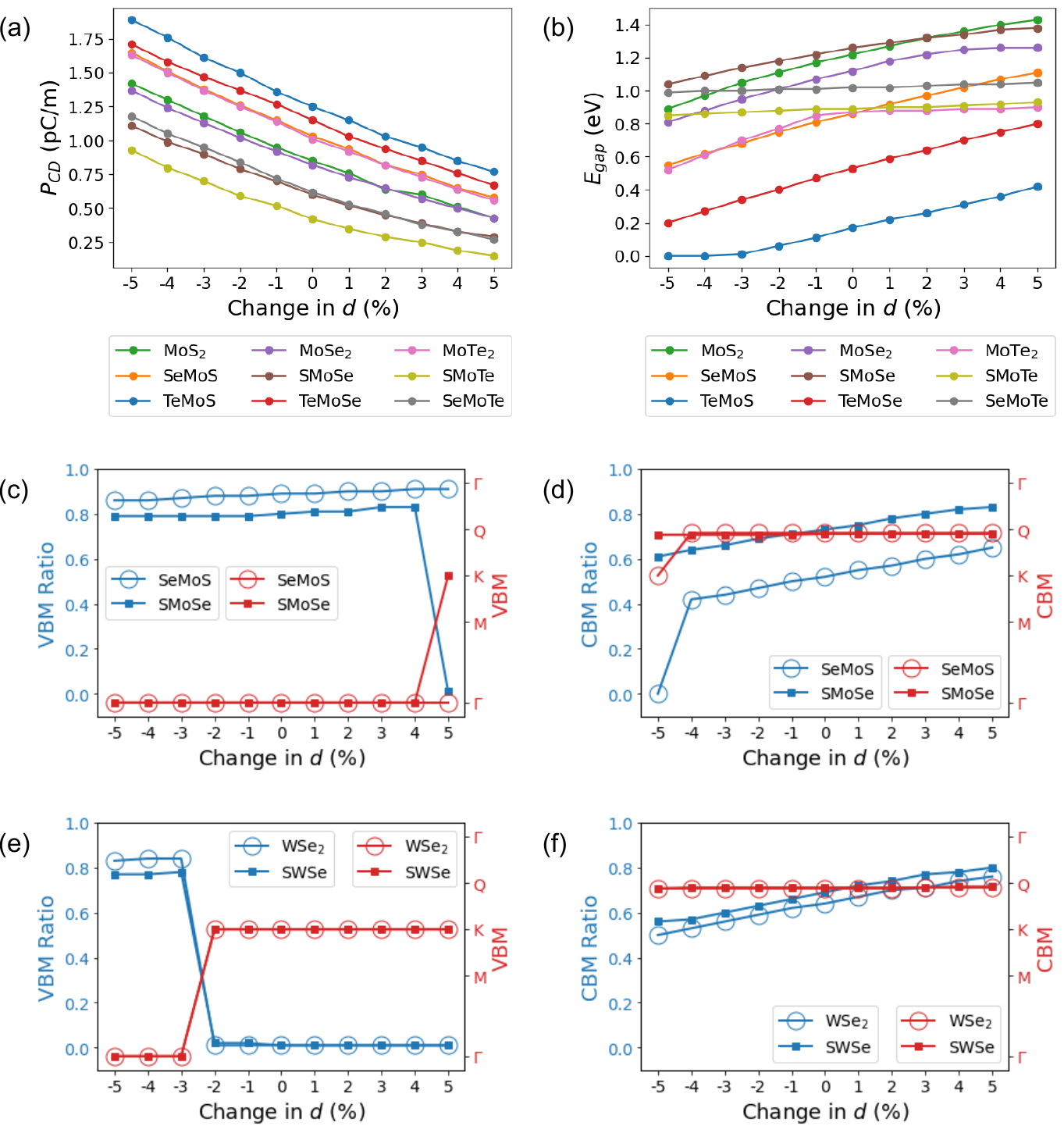}
\caption{\textbf{Polarization and band structure tuning with interlayer distance in bilayer Janus sliding ferroelectrics.} Modulation of (a) $P_{CD}$ (polarization calculated from charge density), and (b) $E_{gap}$ (electronic bandgap) with change in the interlayer distance, $d$, for the XMoY bilayers. Tuning of (c, e) VBM Ratio (and VBM), and (d, f) CBM Ratio (and CBM) with change in $d$ for (c-d) SeMoS and SMoSe bilayers, and (e-f) WSe$_2$ and SWSe bilayers. Here, VBM and CBM stand for valence band maximum and conduction band minimum, respectively.}  
\label{f6}
\end{center}
\end{figure*}
%%%% Figure 6 %%%%

We have identified the tuning of the interlayer distance ($d$) as the mechanism behind the polarization and bandgap modulation in Janus sliding ferroelectrics. A widely used approach to tune the interlayer distance in bilayer and heterostructured TMDs is the application of out-of-plane hydrostatic pressure using a diamond anvil cell~\cite{Qiao2022,Li2023}. In addition, in-plane strain engineering has also been shown to effectively modulate the interlayer coupling~\cite{Pak2017,Yan2020}. In this section, we explicitly calculate the effect of the variation of $d$ on the polarization and band structure of the bilayer sliding ferroelectrics.

Figure~\ref{f6}(a) and Figure~\ref{f6}(b) show how the polarization obtained from charge density ($P_{CD}$) and the electronic bandgap($E_{gap}$) vary with $d$ for the XMoY bilayers. As $d$ decreases, $P_{CD}$ increases, whereas $E_{gap}$ decreases. The variations in $P_{CD}$ and $E_{gap}$ with $d$ are also observed in parent TMD bilayers, further underscoring the crucial role of interlayer vdW interactions in tuning the properties of TMD-based sliding ferroelectrics. Notably, the TeMoS bilayer becomes semi-metallic when $d$ is reduced by more than 4\%, as illustrated in Figure~\ref{f6}(b) and Figure S13~\cite{SuppMat}. A comparison of the percentage changes in $P_{CD}$ and $E_{gap}$ with $d$, shown in Figure S14~\cite{SuppMat}, reveals that E-in stacked Janus bilayers exhibit a greater variation in $E_{gap}$ and a smaller variation in $P_{CD}$ compared to their E-out stacked counterparts.

To further investigate pressure-induced phase transitions and tunability, we compressed the XMoY bilayers up to a 10\% reduction in $d$, with the results shown in Figure S13~\cite{SuppMat}. As expected, additional compression led to further bandgap reduction accompanied by enhanced polarization, while the TeMoS bilayer remained semimetallic with a zero bandgap. Notably, at 8\% and 9\% reduction in $d$, both TeMoSe and MoTe$_2$ also displayed zero bandgaps and semimetallic behavior, as confirmed by their band structures. These observations reinforce the possibility of pressure-induced transitions from semiconducting to semimetallic or metallic states in TMD bilayers.

Figure~\ref{f6}(c) and Figure~\ref{f6}(d) display the variation of the VBM and CBM ratios, respectively, with $d$ for SeMoS (E-in stacking) and SMoSe (E-out stacking) bilayers. Both ratios show a slight overall increase with increasing $d$, with more pronounced changes in the CBM ratios. These variations are generally monotonic, except at points where the VBM or CBM undergoes a band-edge shift. For example, a shift in the VBM is observed at a 5\% increase in $d$ for the SMoSe bilayer (Figure~\ref{f6}(c)), and a shift in the CBM occurs at a 5\% decrease in $d$ for the SeMoS bilayer (Figure~\ref{f6}(d)). Figure S15 and Figure S16\cite{SuppMat} present similar trends in the VBM and CBM ratios across all XMoY bilayers, further confirming the correlation between interlayer distance and band-edge contributions. 

As shown in Figure S15 (c)~\cite{SuppMat}, a notable shift of the VBM from the K point to the $\Gamma$ point occurs in the MoTe$_2$ bilayer with just a 1\% decrease in $d$. This shift leads to a sudden increase in the VBM ratio, indicating a reduced depolarization effect from hole doping when $d$ is decreased beyond 1\%. Interestingly, a similar behavior is observed for MoTe$_2$, WSe$_2$, and SWSe bilayers when SOC is turned off: the VBM shifts and the VBM ratio increase, as shown in Figure S8~\cite{SuppMat}, which results in a reduction in hole-doping-induced depolarization, as shown in Figure S9-S10~\cite{SuppMat}. These observations suggest that decreasing the interlayer distance could likewise enhance the VBM ratio and reduce hole-doping-induced depolarization in WSe$_2$ and SWSe bilayers.        

To validate our intuition regarding the enhancement of the VBM ratio with reduced interlayer distance ($d$) in WSe$_2$ and SWSe bilayers, we analyze the variation of VBM and CBM ratios with $d$, as shown in Figure~\ref{f6} (e)-(f). Both WSe$_2$ and SWSe bilayers exhibit a sudden increase in the VBM ratio when $d$ is decreased beyond 3\%, which corresponds to a shift of the VBM from the K point to the $\Gamma$ point, as shown in Figure~\ref{f6}(e). In contrast, the CBM ratio increases monotonically with increasing $d$, as shown in Figure~\ref{f6} (f), similar to the behavior observed in the XMoY bilayers. The enhanced VBM ratios for WSe$_2$ and SWSe bilayers closely resemble those obtained when the SOC effect is switched off, as shown in Figure S8~\cite{SuppMat}. Therefore, we anticipate a comparable reduction in hole-doping-induced depolarization in WSe$_2$ and SWSe bilayers for reduced $d$, consistent with the effect observed upon disabling SOC (Figure S10~\cite{SuppMat}).

\subsection*{Trilayer Janus Sliding Ferroelectrics}

\label{tri}
%%%% Figure 7 %%%%
\begin{figure*}[t]
\begin{center}
\includegraphics[width=0.75\linewidth]{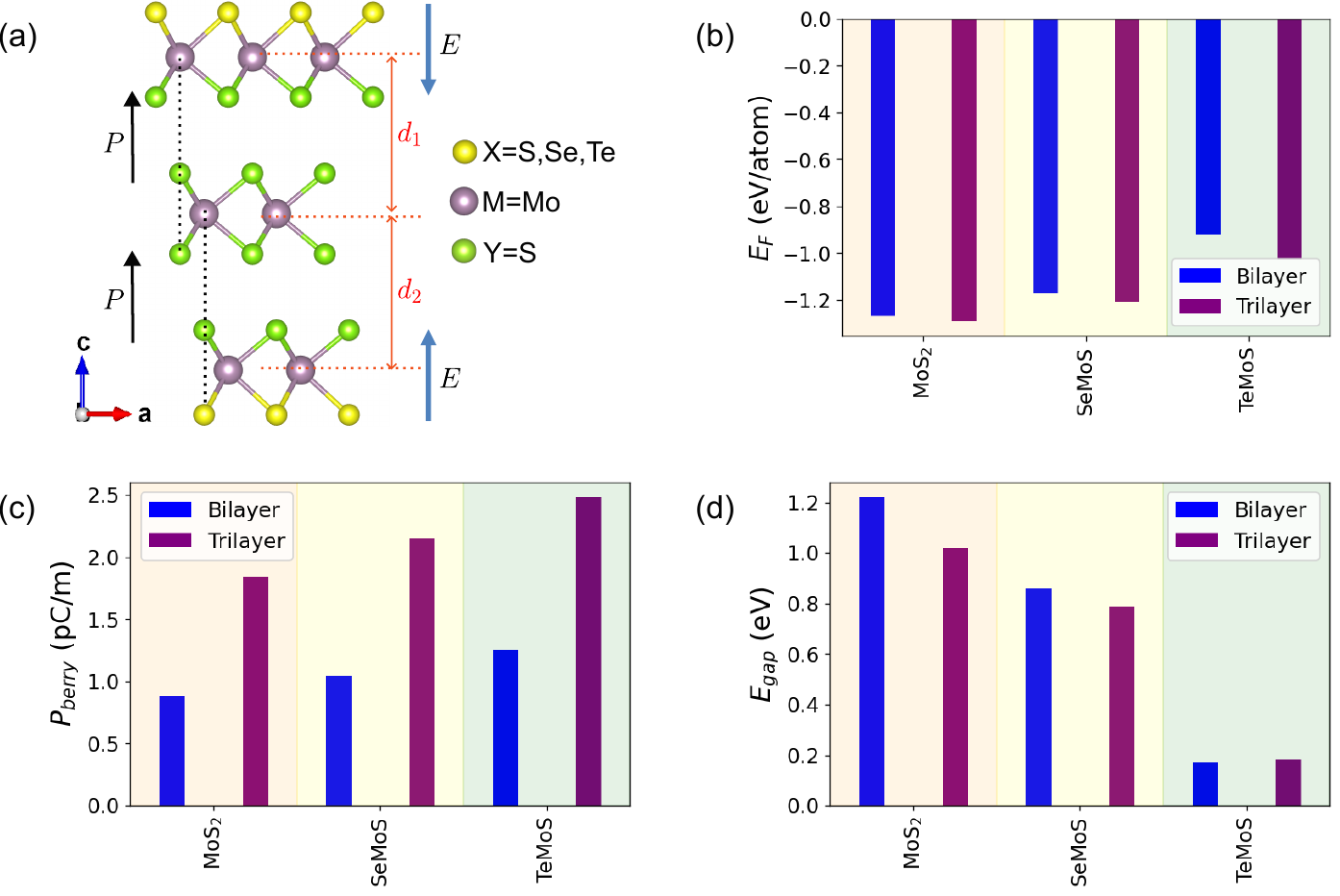}
\caption{\textbf{Crystal structure of trilayer Janus sliding ferroelectrics and comparison with their bilayer counterparts.} (a) Side view of the proposed ABC stacked trilayer sliding ferroelectric XMoS structure. Here, $d_1$ and $d_2$ represent the interlayer distances. The direction of polarization is depicted by the black arrows. The blue arrows represent the direction of the intrinsic electric field, E, in the case of Janus structures with X$\neq$Y. The dashed black line allows for better visualization of the ABC stacking. Here, $a$ is along the in-plane $x$-axis and $c$ is along the stacking direction, i.e., along the $z$-axis. (b) Formation energy (E$_F$), (c) polarization via Berry phase method ($P_{berry}$), and (d) electronic bandgap ($E_{gap}$) for the bilayer and trilayer of MoS$_2$, SeMoS, and TeMoS.}  
\label{f7}
\end{center}
\end{figure*}
%%%% Figure 7 %%%%

\subsubsection*{\textbf{Structure and Stability}}
\label{structri}

Motivated by the potential for enhanced polarization with multiple stable polarization states~\cite{Deb2022,Cao2024}, and the possibility of increased interlayer vdW interactions when only one Janus monolayer is present at the interface~\cite{Shengxi2020,Shengxi2021}, we propose an ABC-stacked trilayer structure. In this design, a MoS$_2$ monolayer is sandwiched between two Janus monolayers, with the intrinsic electric fields ($E$) in the Janus layers oriented toward the heterostructure interface, as illustrated in Figure~\ref{f7}(a).

The ABC stacking configuration shown in Figure~\ref{f7}(a) closely resembles the AB stacking illustrated in Figure~\ref{f1}(a), producing an upward out-of-plane polarization at each interface. The interlayer distances at the top and bottom interfaces, denoted as $d_1$ and $d_2$, respectively, are key factors in modulating the vdW interactions and the resulting interfacial polarizations. As shown in Table S4~\cite{SuppMat}, $d_1$ and $d_2$ have nearly identical values, indicating similar interlayer interactions and, consequently, comparable polarization at both interfaces, which is consistent with the behavior expected from interfacial ferroelectrics.   

To realize Janus sliding ferroelectrics with ABC stacking, we replaced the outermost S layers with Se or Te. This modification generates an intrinsic electric field ($E$) pointing inward at each interface, as illustrated in Figure~\ref{f7}(a), resembling the E-in stacking configuration of Janus bilayers shown in Figure~\ref{f1}(b).

The trilayer systems have lower (more negative) formation energies than their corresponding bilayer systems, indicating enhanced stability, as shown in Figure~\ref{f7}(b). We further validated the dynamical stability of both the Janus trilayers and their bilayer counterparts by calculating their phonon dispersion curves (Figure S18~\cite{SuppMat}), which show no imaginary frequencies. Additionally, the thermal stability of the Janus trilayers and their bilayer counterparts was analyzed using AIMD simulations at 300 K and 600 K, as shown in the section \textit{Thermal Stability Analysis} of the Supplementary Material~\cite{SuppMat}. No persistent structural changes were observed during the energy evolution, indicating robust thermal stability for these systems. 

Although this work focuses on Janus trilayer systems with $E$ fields directed toward the interface, the same design strategy can be extended to configurations where the $E$ fields point outward, away from the interface (toward the vacuum). 

\subsubsection*{\textbf{Polarization Modulation}}
\label{poltri}

In Figure~\ref{f7}(c), we compare the polarization values for the trilayer systems with those of their bilayer counterparts. Interestingly, similar to the bilayers, the trilayer Janus sliding ferroelectrics exhibit higher polarization when the $E$ fields point toward the interface. These $E$ fields also lead to a reduction in the interlayer distances, $d_1$ and $d_2$, as shown in Table S4~\cite{SuppMat}. Additionally, the charge density profiles of the Janus trilayers, shown in Figure S19~\cite{SuppMat}, display similar modulation due to the $E$ fields as observed in the E-in stacked Janus bilayers. These results suggest that, in ABC-stacked Janus trialyers, the intrinsic $E$ fields modulate the polarization by redistributing interfacial electron density and tuning the interlayer spacing\textemdash mirroring the mechanism found in AB-stacked Janus bilayers.

The potential profiles in Figure S19~\cite{SuppMat} confirm the interfacial nature of the polarization in the trilayer systems, which results in the cumulative nature of polarization at each interface. This explains the higher polarization values observed in Figure~\ref{f7}(c) compared to their bilayer counterparts. Additionally, the $\Delta\phi$ and $P_{CD}$ values follow a similar trend, further supporting the observed polarization behavior, as shown in Table S4~\cite{SuppMat}. 

\subsubsection*{\textbf{Band Structure Modulation}}
\label{bandtri}

A comparison of the electronic bandgap values for the trilayer systems and their bilayer counterparts in Figure~\ref{f7}(d) reveals a similar bandgap reduction driven by the intrinsic electric fields ($E$) within the Janus monolayers, consistent with the behavior observed in E-in stacked Janus bilayers. Notably, the trilayer systems exhibit slightly lower bandgap values than the bilayers. Since both polarization and bandgap are modulated by interlayer distance ($d$) in Janus bilayers, we attribute the bandgap tuning in Janus trilayers to the same mechanism, as discussed for polarization modulation.   

The band structures of the trilayer systems exhibit strong asymmetry in the layer-wise contributions to the VBM and CBM. This leads to an uneven distribution of the partial charge density associated with the VBM and CBM, as illustrated in Figure S20~\cite{SuppMat}, with the asymmetry becoming more pronounced in the Janus sliding ferroelectric configurations. Consequently, based on the earlier discussion of doping-induced depolarization in Janus bilayers, we anticipate a significantly stronger depolarization effect from charge doping in trilayer systems compared to their bilayer counterparts.

\section*{Conclusions and outlook}
\label{conclusion}

In conclusion, our in-depth analysis of the TMD-based AB-stacked bilayer Janus sliding ferroelectrics reveals that the direction of the intrinsic electric field within the Janus monolayers can significantly modulate the polarization and the band structures of the bilayer sliding ferroelectrics. The polarization and the bandgap modulation are predominantly driven by the modulation of the interlayer distance, which is achieved by the redistribution of the accumulated interlayer charge via the intrinsic electric fields of the Janus monolayers. In addition to the bandgap, the intrinsic electric fields of the Janus monolayers can also modulate the layer-wise contribution of the valence and conduction bands in the Janus sliding ferroelectrics. This results in the modulation of the depolarization caused by the extrinsic charge dopants and thus suggests a new mechanism to modulate the unique existence of metallicity with ferroelectric polarization in TMD-based sliding ferroelectrics. In addition, we analyzed the effect of the variation of the interlayer distance on the tuning of the properties of the TMD-based sliding ferroelectric bilayers. In the end, we proposed a material design for a similar modulation of polarization and electronic bandgap in the ABC-stacked trilayer Janus sliding ferroelectrics, expanding the playground for designing new sliding ferroelectrics. Such tunability of polarization and band structure provides an opportunity to design materials for specific logic and memory devices and also energy harvesting applications, in the future.

Finally, we would like to suggest some interesting future avenues for experimental and theoretical work motivated by our results. Experiments confirming our theoretical finding of enhancement of polarization along with bandgap and hole-doping-induced depolarization reduction by out-of-plane compressive strain for AB-stacked WSe$_2$ bilayer would be an interesting and useful finding, considering the already existing experimental work on the doping-induced polarization modulation for WSe$_2$ bilayer in Ref.~\cite{Deb2022}. Also, the semi-metallic nature observed after reducing the interlayer distance for the TeMoS bilayer suggests the possibility of designing Janus sliding ferroelectrics with intrinsic semi-metallicity by stacking them in E-in stacking, given much lower bandgaps for parent TMD bilayers. We expect this work to motivate several other theoretical and experimental investigations to design better-performing sliding ferroelectrics with additional properties such as a low bandgap and polar metallicity.

\subsection*{Supplementary Material}

Formation energy plot for bilayers; charge density and potential profiles for AB and BA stacking of SeMoS bilayer; charge density and potential profiles for XMoY bilayers; correlation plots for $\Delta\phi$ and $P_{CD}$ with $P_{berry}$ along with comparison of SOC and without SOC results; additional results for XMoY bilayers with $d=d_{avg}$; band structures of XMoY bilayers; comparison of band structure of MoTe$_2$, WSe$_2$ and SWSe bilayer without and with SOC; effect of doping on polarization for bilayers; variation of Fermi energy and DOS with doping for TeMoS and SMoTe bilayers; polarization and bandgap variation with more than 5\% reduction in interlayer distance for XMoY bilayers along with band structures for bilayers showing semi-metallic nature at reduced interlayer distance; comparison of percentage change in bandgap and polarization with change in $d$ for XMoY bilayers; variation of the VBM and CBM ratios with change in $d$ for XMoY bilayers; variation of polarization and bandgap with change in $d$ for WSe$_2$ and SWSe bilayers; phonon dispersion curves; charge density and potential profiles for trilayers; band structure with layer-wise contribution analysis for trilayers; tables for data for bilayer and trilayer systems presented and referred in the paper; methodology for $P_{CD}$ calculation; thermal stability analysis (PDF).

\subsection*{Acknowledgements}
We thank M. Jain and H. R. Krishnamurthy for stimulating discussions and invaluable suggestions. AM also thanks Nesta Benno Joseph and Nayana Devaraj for their constructive feedback on the manuscript. AM acknowledges support from the Prime Minister's Research Fellowship (PMRF). AN acknowledges support from Anusandhan National Research Foundation (ANRF) through grant number CRG/2023/000114.

\bibliography{ref}

\end{document}

%\begin{titlepage}
\title{Supplementary Material:\\
Modulation of Polarization and  Metallicity in Janus Sliding Ferroelectrics}

\author{Akshay Mahajan} \email[]{akshaymahaja@iisc.ac.in} \affiliation{Solid State and Structural Chemistry Unit,
Indian Institute of Science, Bangalore 560012, India.}
\author{Awadhesh Narayan} \email[]{awadhesh@iisc.ac.in} \affiliation{Solid State and Structural Chemistry Unit,
Indian Institute of Science, Bangalore 560012, India.}

\vskip 0.25cm

\date{\today}

%\pacs{}
\maketitle

%\end{titlepage}

\begin{figure}[!htb]
\includegraphics[width=0.57\linewidth]{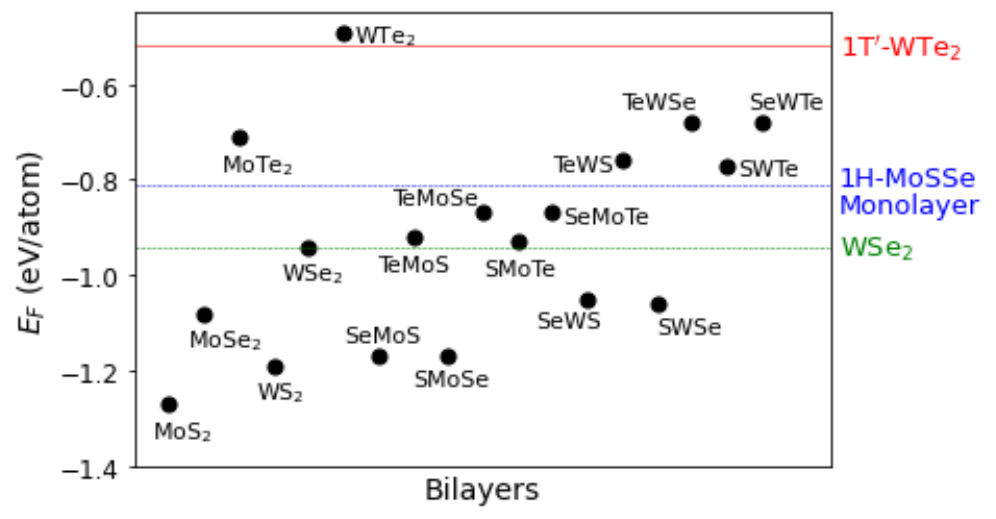}
\caption{Formation energies ($E_F$) for all the sliding ferroelectric bilayers XMY. The red, blue, and green lines denote the formation energy of the 1T$'$-WTe$_2$ bilayer (-0.52 eV/atom), 1H-MoSSe monolayer (-0.81 eV/atom), and the AB-stacked WSe$_2$ bilayer(-0.94 eV/atom), respectively.}       
\label{fs1}
\end{figure}

\begin{figure}[!htb]
\includegraphics[width=0.6\linewidth]{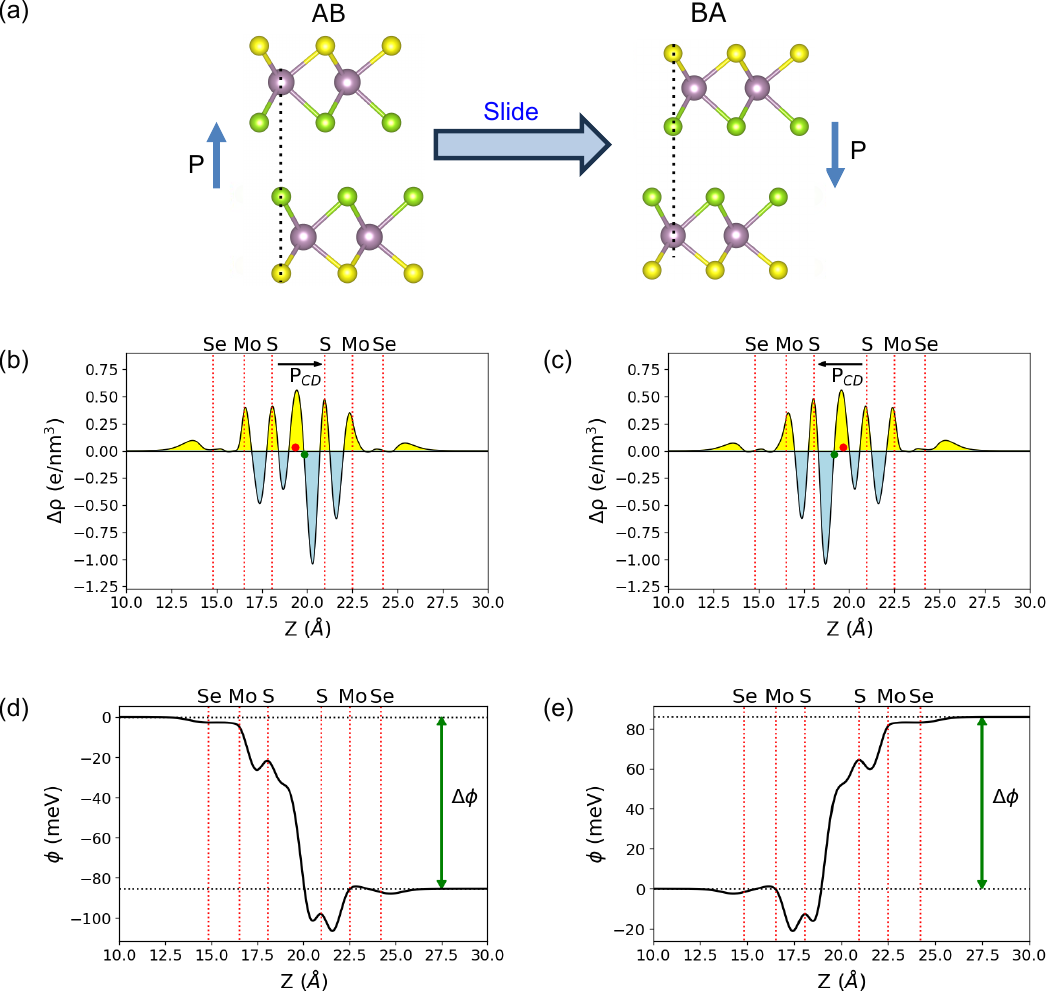}
\caption{(a)Switching the direction of the out-of-plane polarization by switching from the AB to BA stacking via in-plane sliding motion. The dashed black lines allow for better visualization of the AB and BA stacking. The blue arrows show the direction of polarization in each stacking. (b, c)Charge density and (d, e)potential profiles for SeMoS (b, d) AB, and (c, e) BA stacking bilayer. The yellow and blue color in (b) and (c) represents the electron and hole accumulation, respectively. The switching of the direction of polarization calculated from charge density ($P_{CD}$) can be observed in (b) and (c), accompanied by the switching in the relative positions of pseudo-electron (red dot) and pseudo-hole (green dot). The $\Delta\phi$ in (d) and (e) represents the interlayer potential difference. The dashed red lines represent the vertical location of the atoms.} 
\label{fs2}
%\end{center}
\end{figure}

\begin{figure}[!htb]
%\begin{center}
\includegraphics[width=0.9\linewidth]{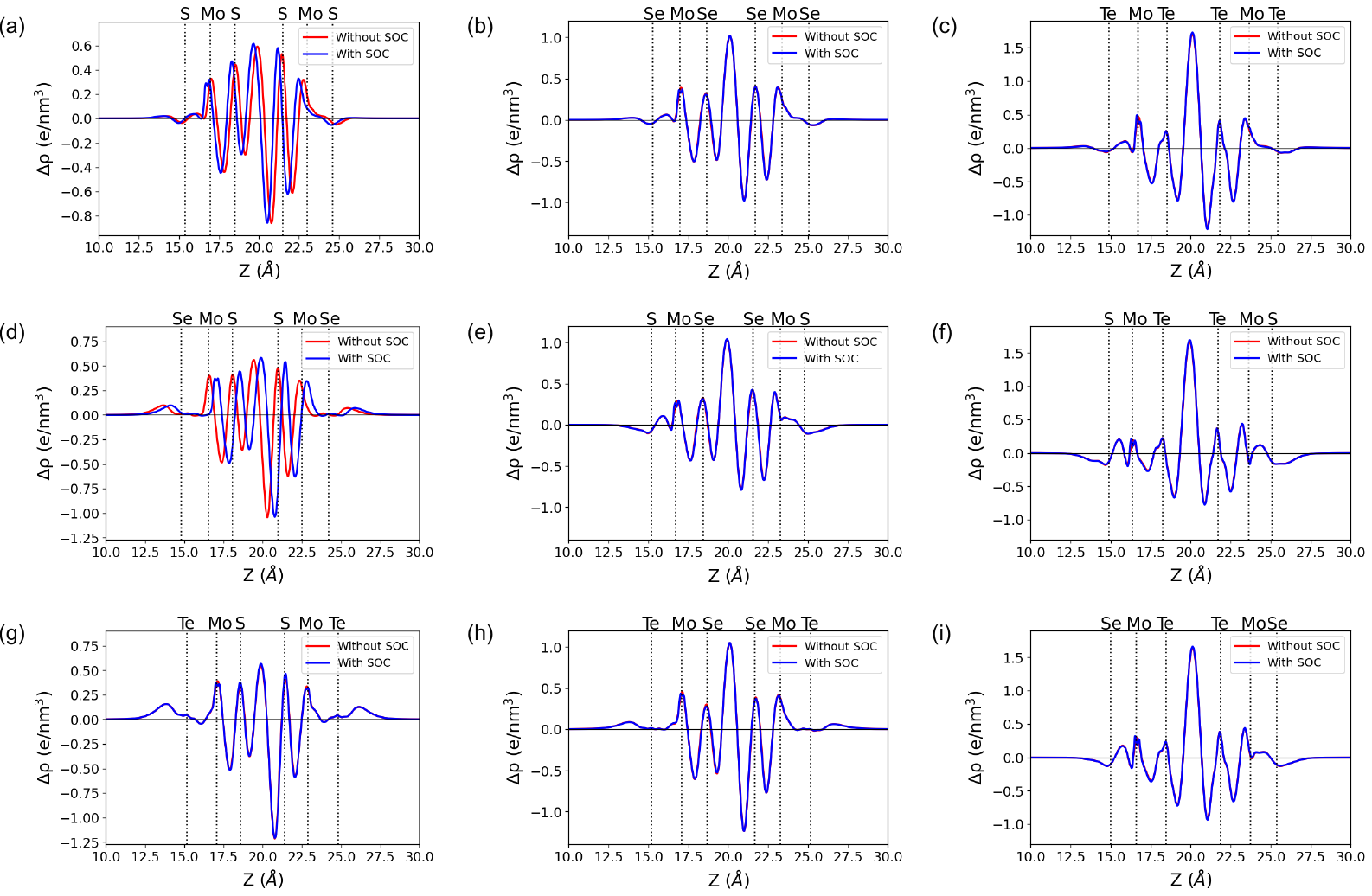}
\caption{Charge density profiles for AB-stacked (a) MoS$_2$, (b) MoSe$_2$, (c) MoTe$_2$, (d) SeMoS, (e) SMoSe, (f) SMoTe, (g) TeMoS, (h) TeMoSe, and (i) SeMoTe bilayer. The red and blue lines represent results without and with spin-orbit coupling (SOC). The dashed black lines represent the vertical location of the atoms.}       
%The yellow and blue color represents the electron and hole accumulation, respectively. The dashed red lines represent the vertical location of the atoms. The direction of polarization calculated from charge density ($P_{CD}$) is from the red dot (pseudo-electron) to the green dot (pseudo-hole), as shown in (a).
\label{fs3}
%\end{center}
\end{figure}

\begin{figure}[!htb]
%\begin{center}
\includegraphics[width=0.9\linewidth]{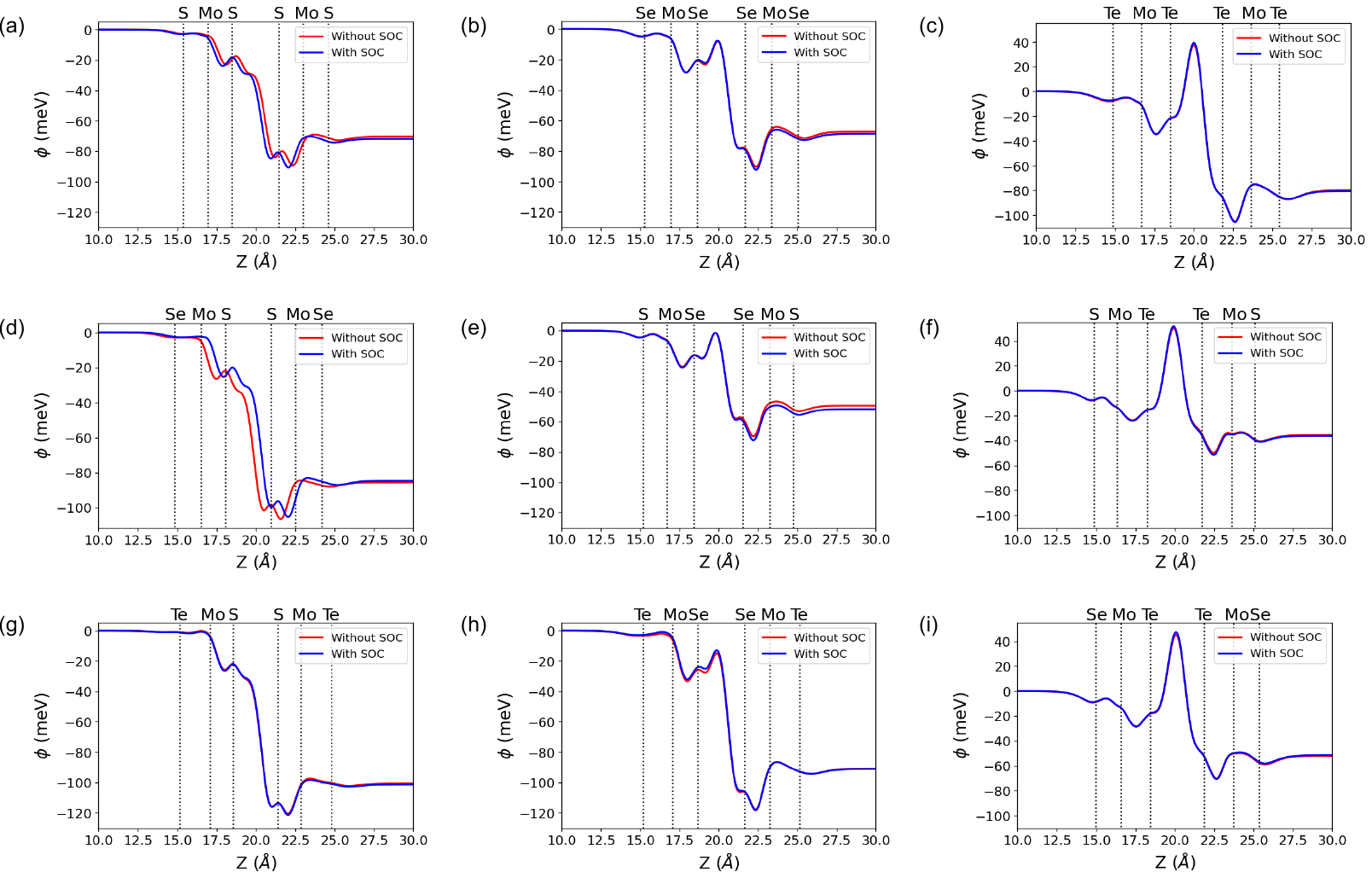}
\caption{Potential profiles for AB-stacked (a) MoS$_2$, (b) MoSe$_2$, (c) MoTe$_2$, (d) SeMoS, (e) SMoSe, (f) SMoTe, (g) TeMoS, (h) TeMoSe, and (i) SeMoTe bilayer. The red and blue lines represent results without and with spin-orbit coupling (SOC). The dashed black lines represent the vertical location of the atoms.}
% The $\Delta\phi$ represents the interlayer potential difference. The dashed red lines represent the vertical location of the atoms.
\label{fs4}
%\end{center}
\end{figure}

\begin{figure}[!htb]
%\begin{center}
\includegraphics[width=0.7\linewidth]{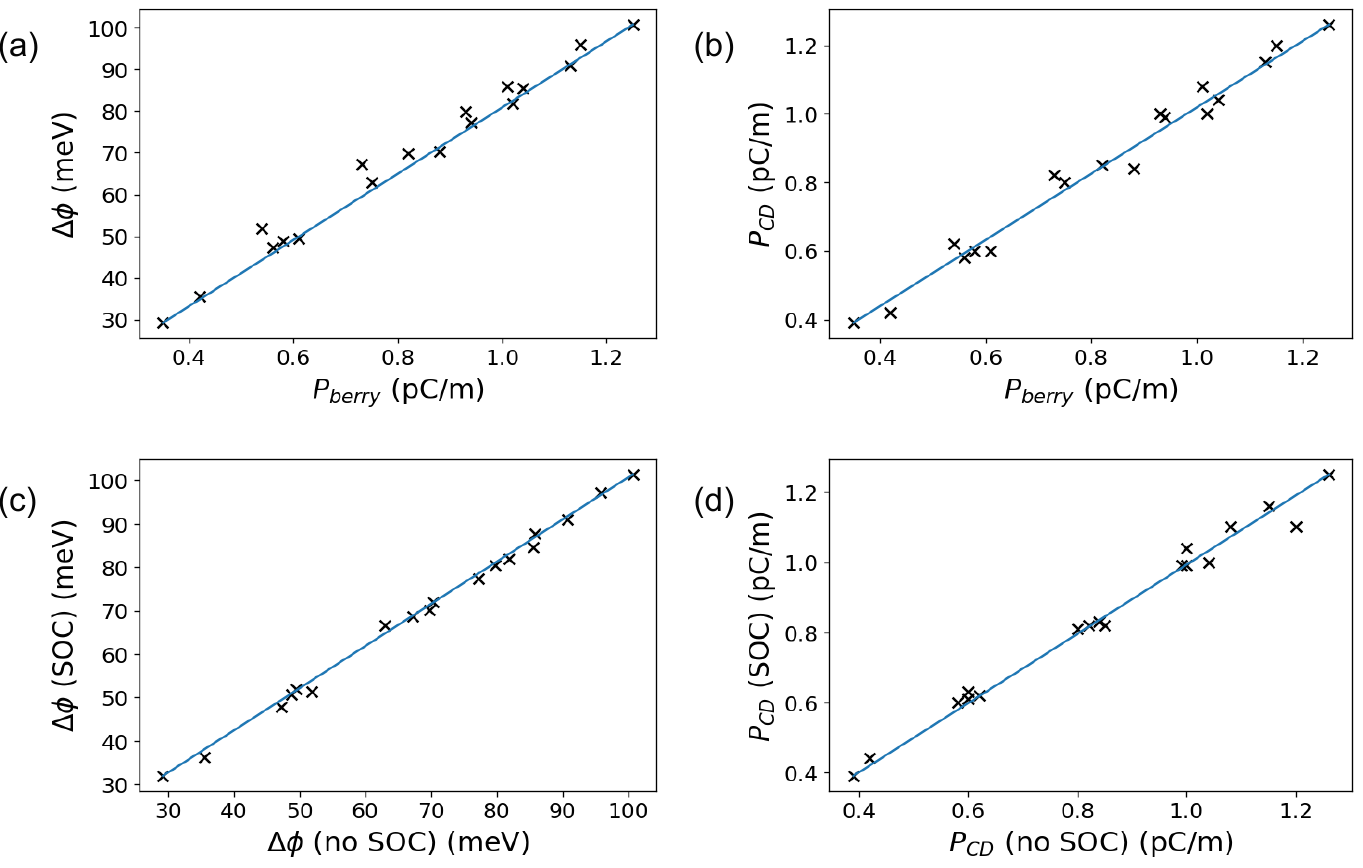}
\caption{Correlation plots for (a) $\Delta\phi$ versus $P_{berry}$, (b) $P_{CD}$ versus $P_{berry}$, (c) $\Delta\phi$ (with SOC) versus $\Delta\phi$ (without SOC), and (d) $P_{CD}$ (with SOC) versus $P_{CD}$ (without SOC), where $\Delta\phi$, $P_{CD}$, $P_{berry}$, and SOC stands for the interlayer potential difference, the polarization via charge density, the polarization obtained via the Berry phase method, and the spin-orbit coupling approximation, respectively. The blue line with a positive slope represents a positive correlation.} 
\label{fs5}
%\end{center}
\end{figure}

\begin{figure}[!htb]
%\begin{center}
\includegraphics[width=0.7\linewidth]{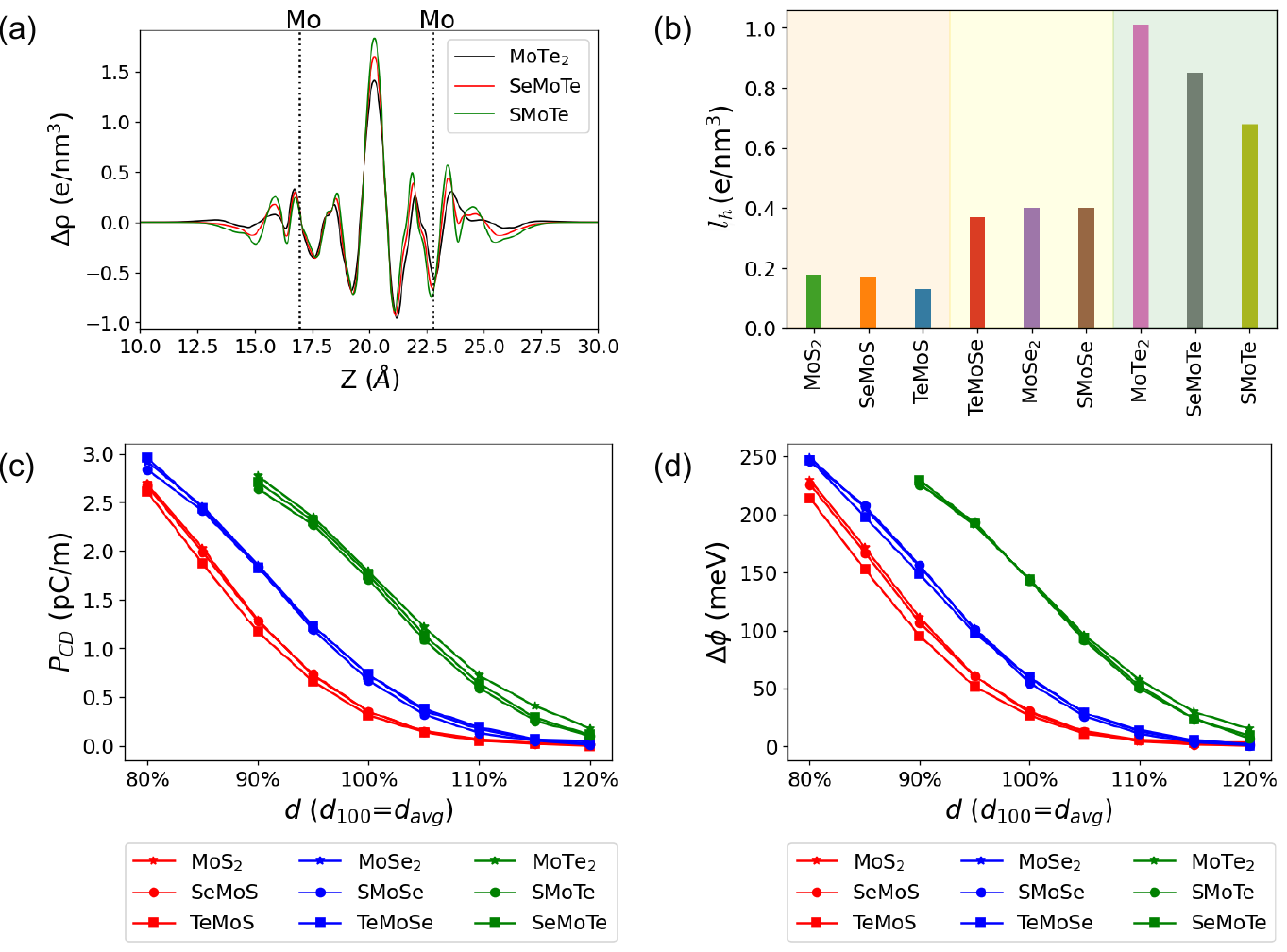}
\caption{(a) Charge density profiles for XMoTe bilayers, and (b) comparison of the parameter $l_h$ of the charge density profiles for XMoY bilayers, with interlayer distance for the XMoY bilayers fixed to the average interlayer distance, $d = d_{avg} = 6.49$\AA. Variation of (c) polarization via charge density ($P_{CD}$), and (d) interlayer potential difference ($\Delta\phi$) for XMoY bilayers with the interlayer distance varying from 80\% to 120\% of the average interlayer distance ($d_{avg}$), with $d_{100}$ representing the interlayer distance at 100\%.}       
\label{fs6}
%\end{center}
\end{figure}

\begin{figure}[!htb]
%\begin{center}
\includegraphics[width=0.9\linewidth]{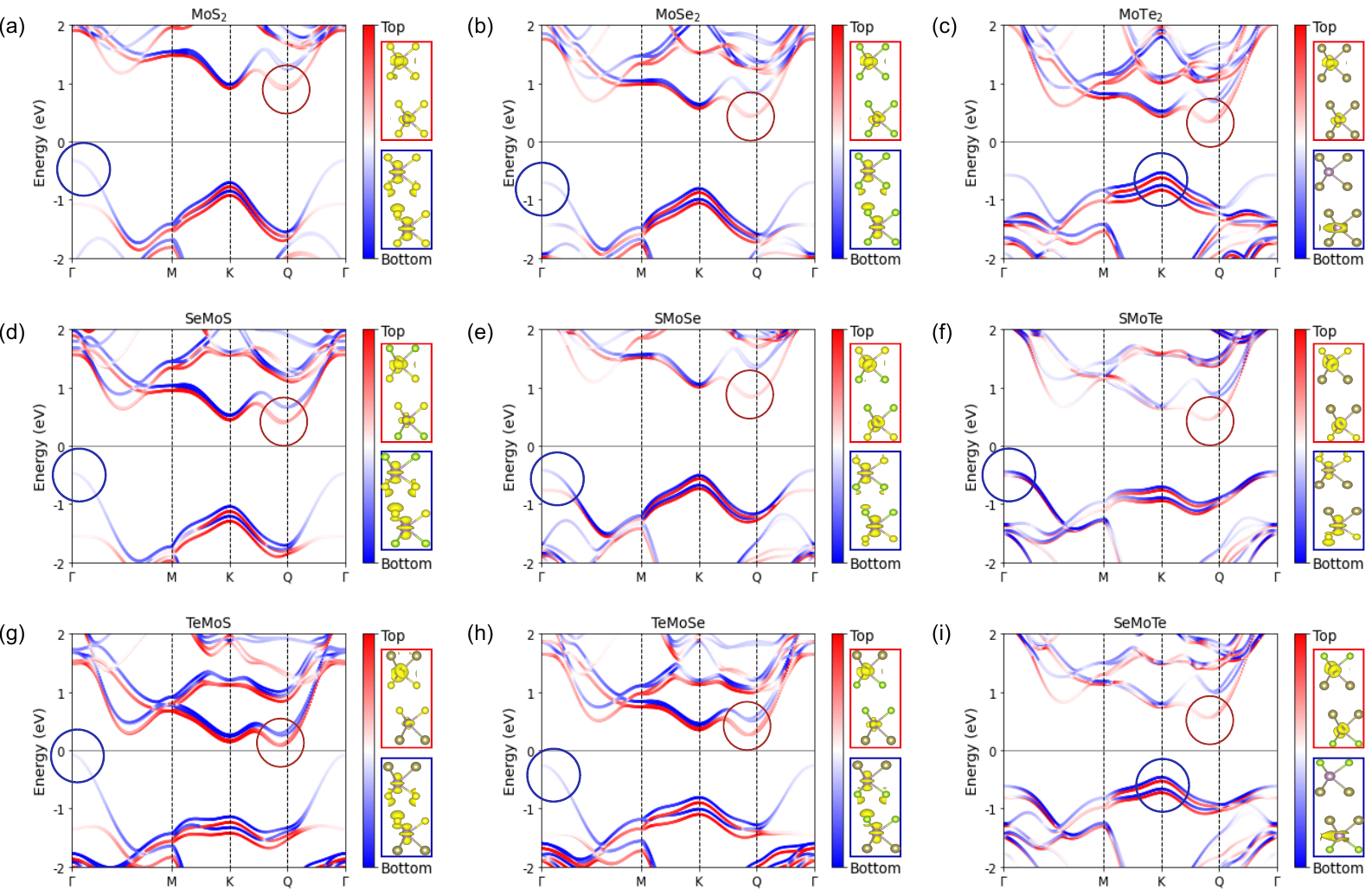}
\caption{Layer-contribution-projected band structure of the XMoY AB-stacked bilayers. The blue and red circles represent the region around the valence band maximum (VBM) and the conduction band minimum (CBM), respectively. The blue and red colors of the bands represent the amount of contribution of the bottom and top layers to the particular band, respectively. Here, Q is the midpoint between K and $\Gamma$. The zero of the energy scale is at the Fermi level. The illustrations on the top right (red color box) and on the bottom right (blue color box) are the partial charge density in the CBM and VBM, respectively.}       
\label{fs7}
%\end{center}
\end{figure}

\begin{figure}[!htb]
\includegraphics[width=0.7\linewidth]{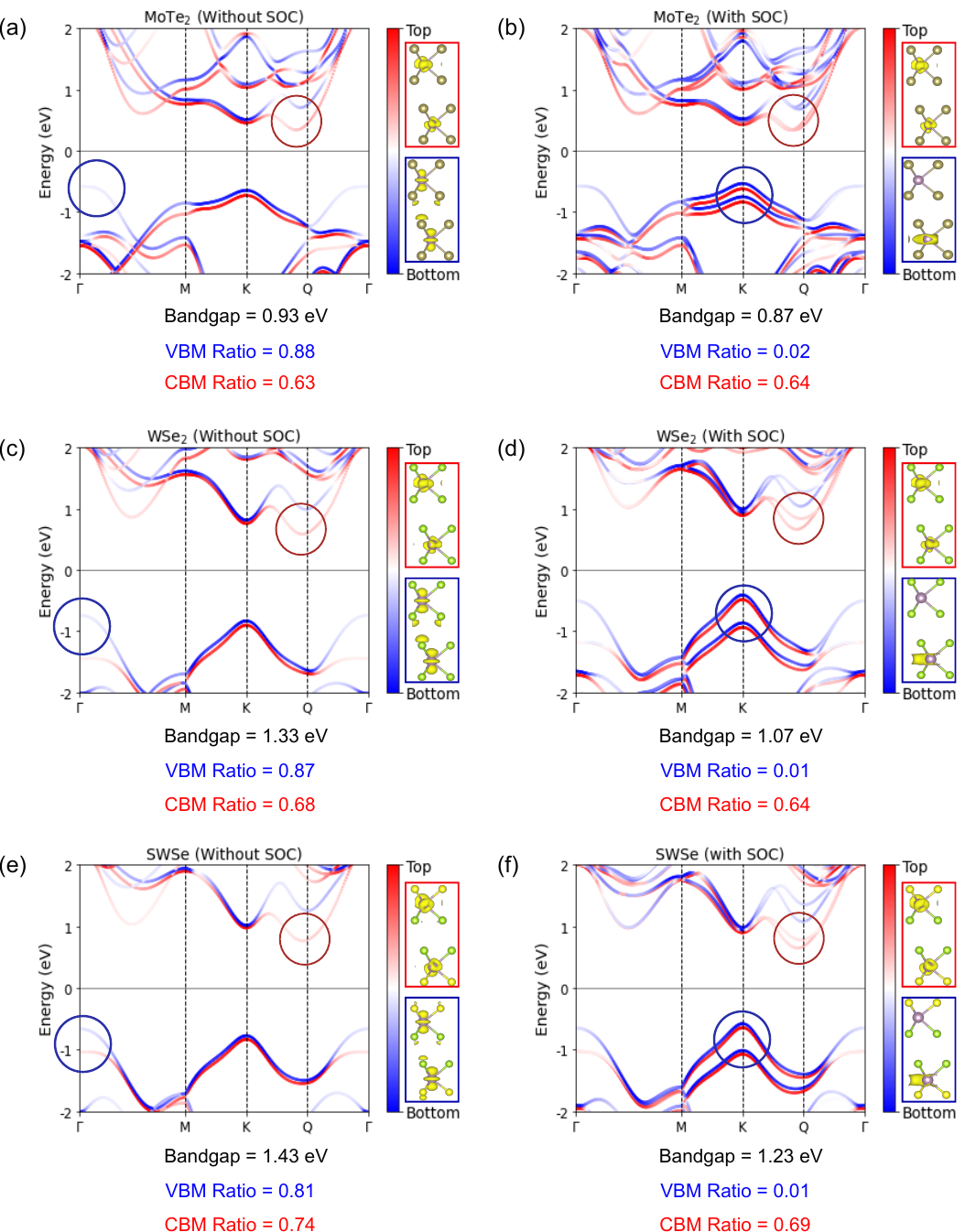}
\caption{Layer-contribution-projected band structure of AB-stacked (a, b) MoTe$_2$, (c, d) WSe$_2$, and (e, f) SWSe bilayers, (a, c, e) without, and (b, d, f) with spin-orbit coupling (SOC). The blue and red circles represent the region around the valence band maximum (VBM) and the conduction band minimum (CBM), respectively. The blue and red colors of the bands represent the amount of contribution of the bottom and top layers to the particular band, respectively. Here, Q is the midpoint between K and $\Gamma$. The zero of the energy scale is at the Fermi level. The illustrations on the top right (red color box) and on the bottom right (blue color box) are the partial charge density in the CBM and VBM, respectively.} 
\label{fs8}
\end{figure}

\begin{figure}[!htb]
%\begin{center}
\includegraphics[width=0.9\linewidth]{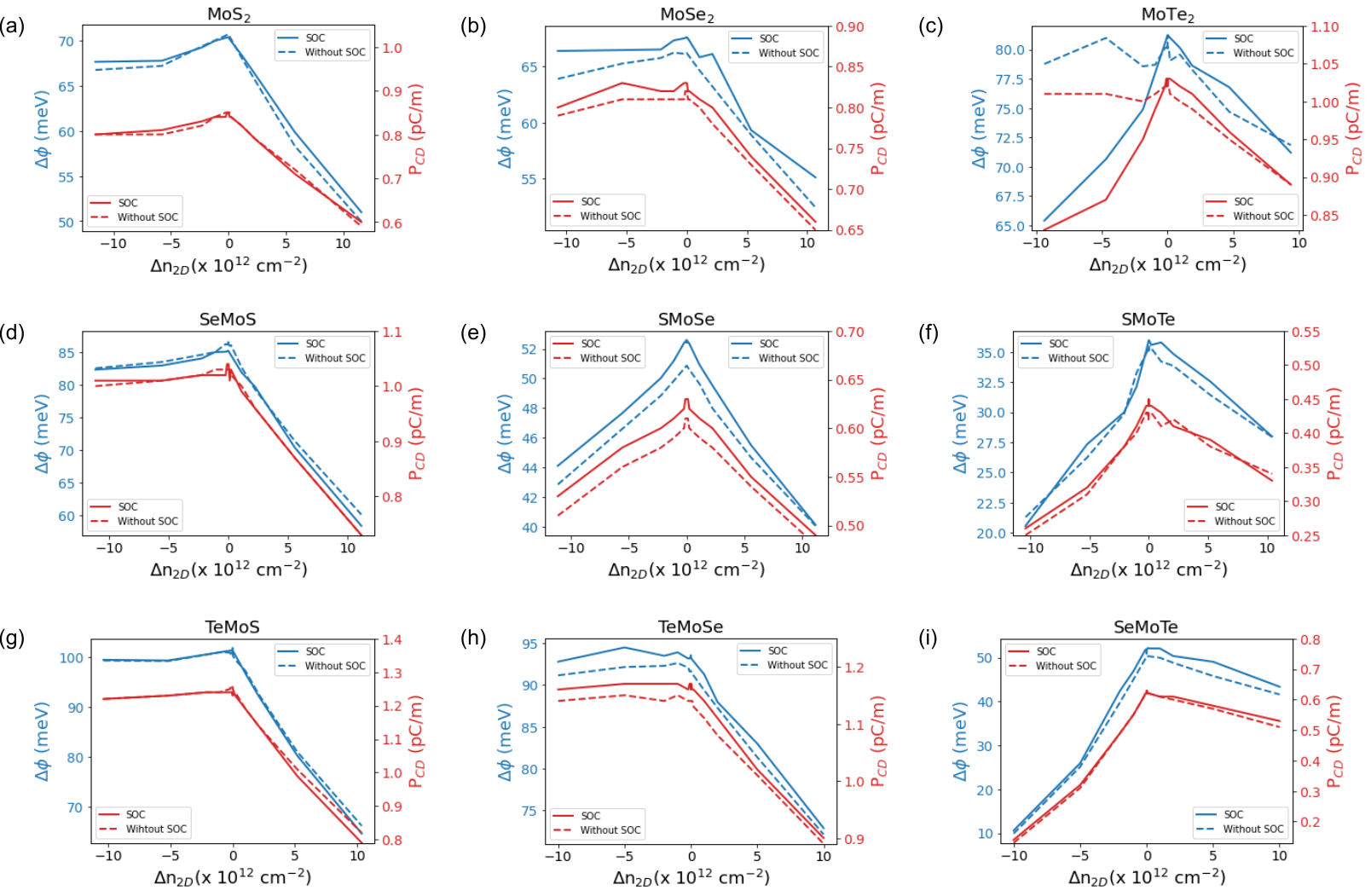}
\caption{Effect of doping ($\Delta$$n_{2D}$) on the interlayer potential difference ($\Delta \phi$, blue color) and polarization via charge density ($P_{CD}$, red color) for the XMoY AB-stacked bilayers. The positive and negative values of $\Delta$$n_{2D}$ correspond to electron and hole doping, respectively. The bold and dashed lines represent results with and without spin-orbit coupling (SOC).}
\label{fs9}
%\end{center}
\end{figure}

\begin{figure}[!htb]
%\begin{center}
\includegraphics[width=0.9\linewidth]{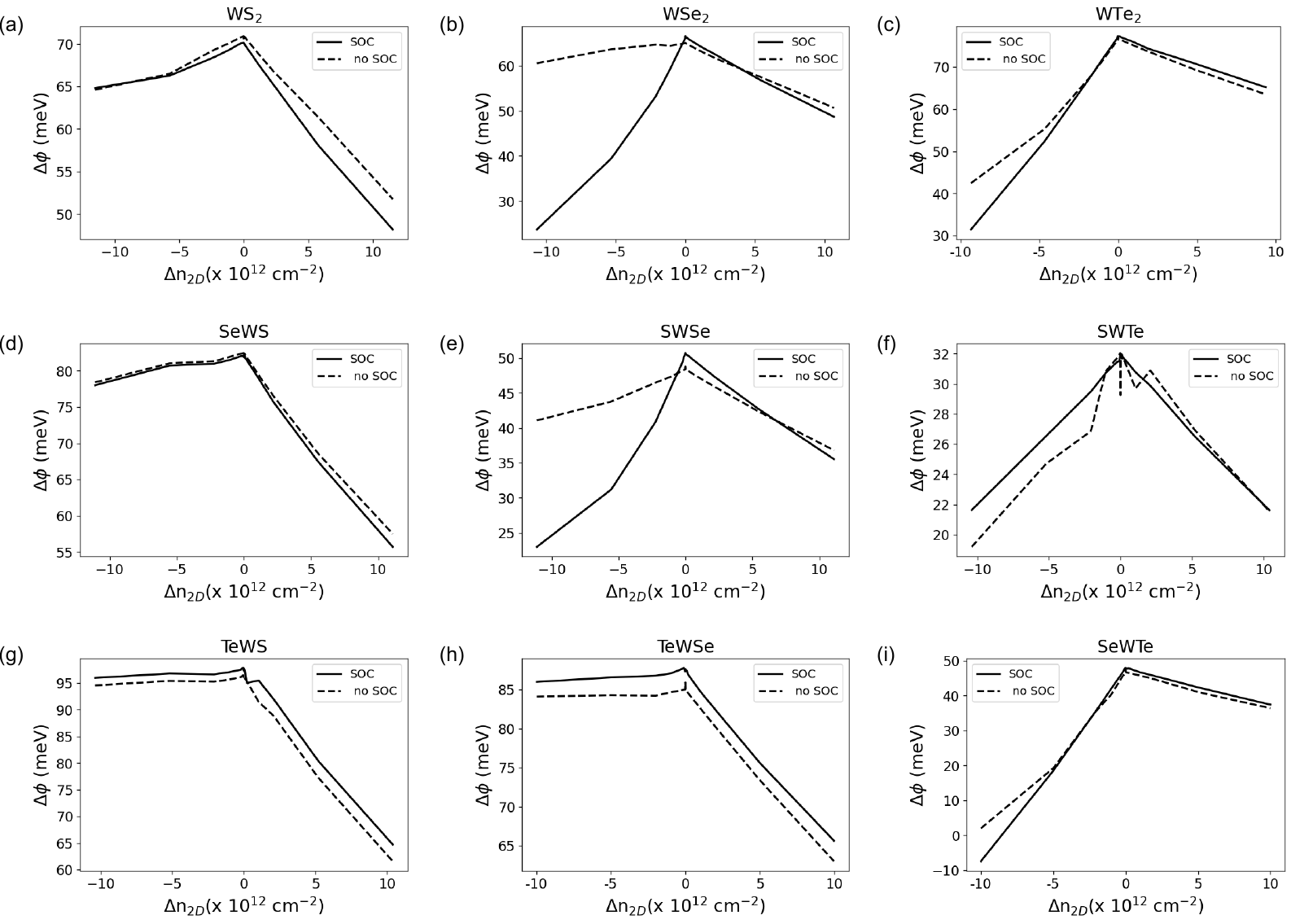}
\caption{Effect of doping ($\Delta$$n_{2D}$) on the interlayer potential difference ($\Delta \phi$) for the XWY AB-stacked bilayers. The positive and negative values of $\Delta$$n_{2D}$ correspond to electron and hole doping, respectively. The bold and dashed lines represent results with and without spin-orbit coupling (SOC).}
\label{fs10}
%\end{center}
\end{figure}

\begin{figure}[!htb]
%\begin{center}
\includegraphics[width=0.7\linewidth]{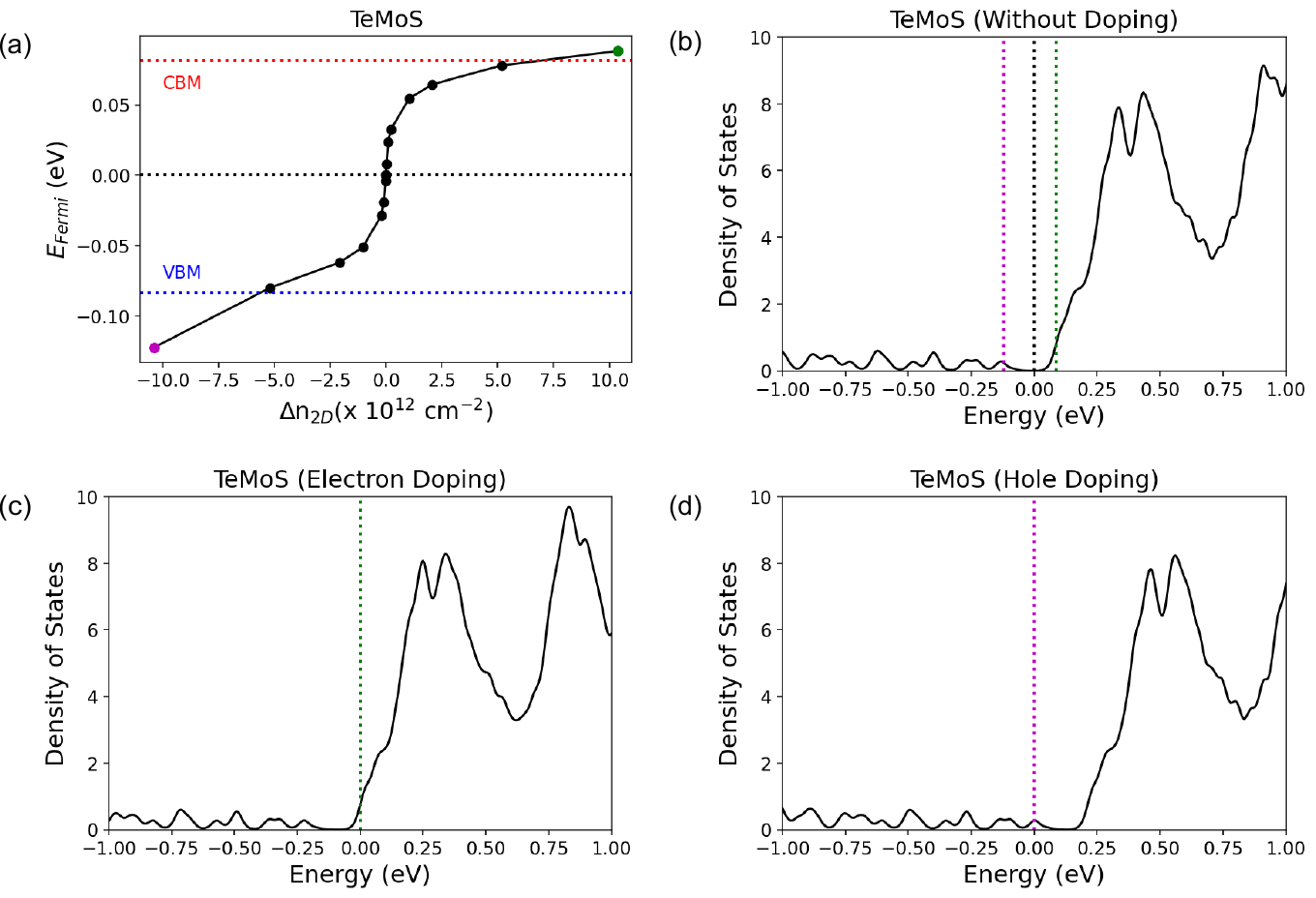}
\caption{(a) Variation of the Fermi energy ($E_{Fermi}$) with doping in the TeMoS bilayer. The dashed black, blue, and red lines represent the Fermi level, valence band maximum (VBM), and conduction band minimum (CBM), respectively, for the undoped case. Density of states for (b) undoped, (c) highest electron-doped, and (d) highest hole-doped TeMoS bilayers, with a maximum doping concentration of $10.38 \times 10^{12}$ cm$^{-2}$. The zero of energy in the DOS plots corresponds to the Fermi level of the respective system. Green and magenta dots and dashed lines indicate the Fermi energy for the highest electron and hole doping, respectively.}
\label{fs11}
%\end{center}
\end{figure}

\begin{figure}[!htb]
%\begin{center}
\includegraphics[width=0.7\linewidth]{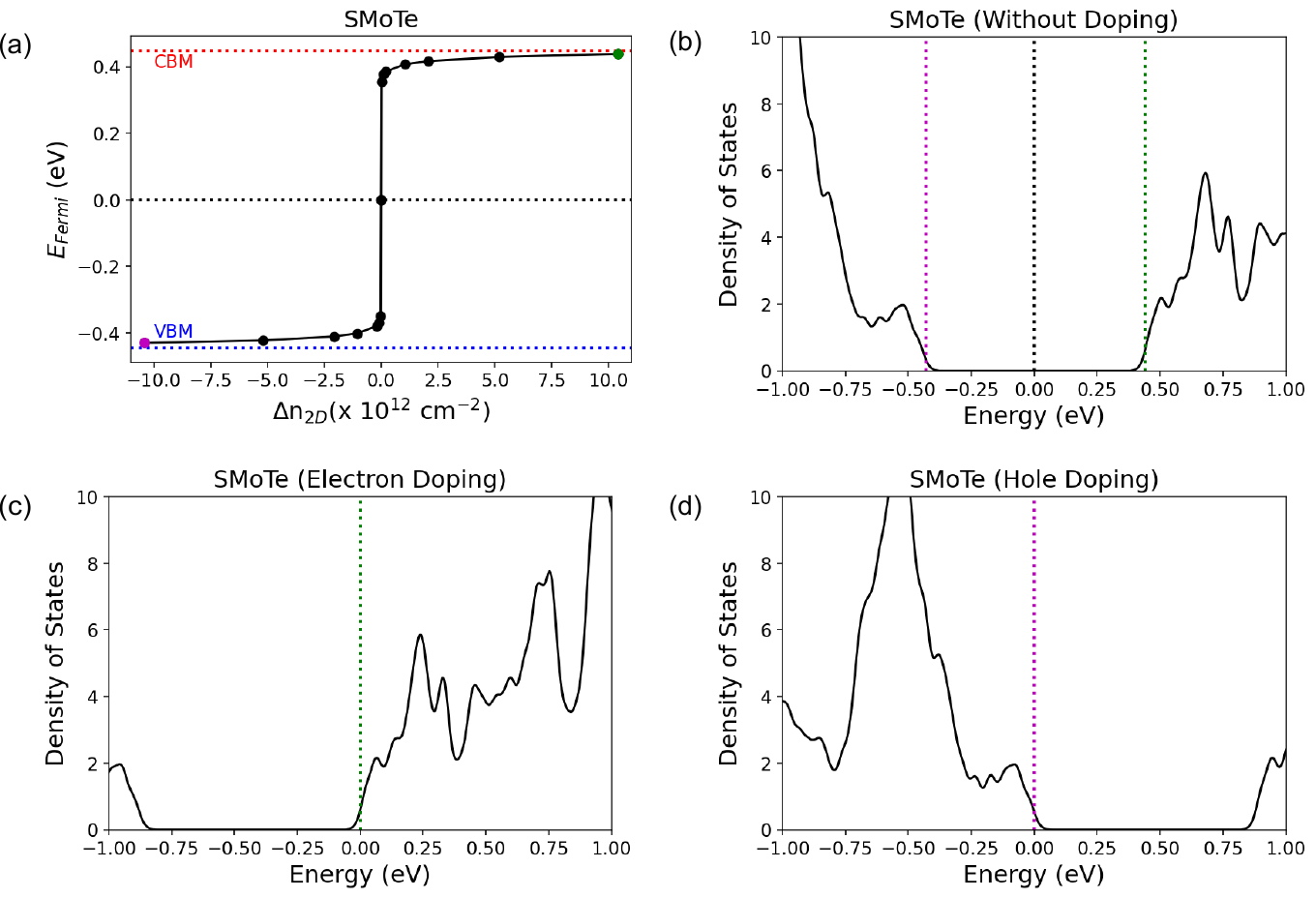}
\caption{(a) Variation of the Fermi energy ($E_{Fermi}$) with doping in the SMoTe bilayer. The dashed black, blue, and red lines represent the Fermi level, valence band maximum (VBM), and conduction band minimum (CBM), respectively, for the undoped case. Density of states for (b) undoped, (c) highest electron-doped, and (d) highest hole-doped TeMoS bilayers, with a maximum doping concentration of $10.38 \times 10^{12}$ cm$^{-2}$. The zero of energy in the DOS plots corresponds to the Fermi level of the respective system. Green and magenta dots and dashed lines indicate the Fermi energy for the highest electron and hole doping, respectively.}
\label{fs12}
%\end{center}
\end{figure}

\begin{figure}[!htb]
%\begin{center}
\includegraphics[width=0.7\linewidth]{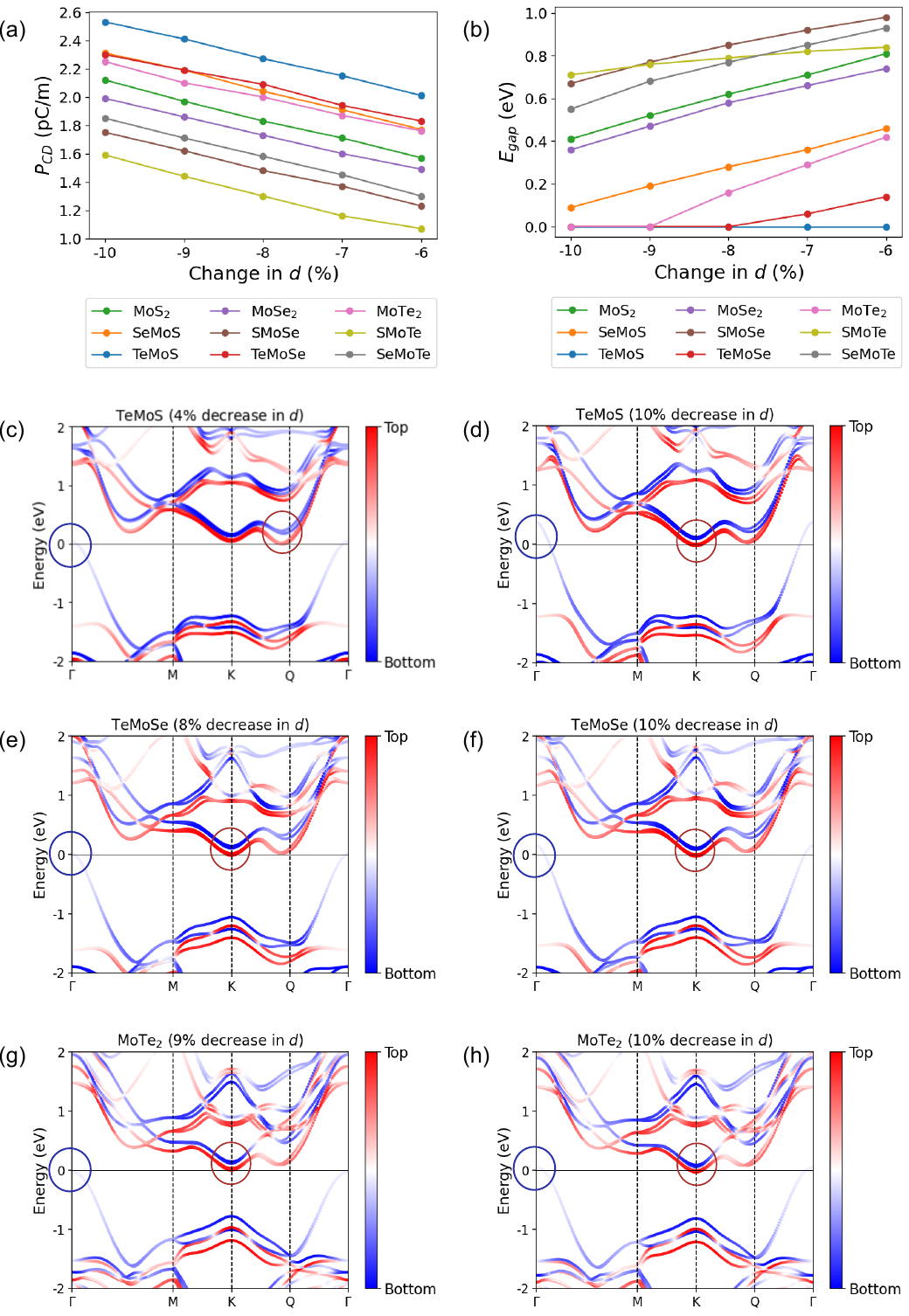}
\caption{Modulation of (a) $P_{CD}$ (polarization calculated from charge density), and (b) $E_{gap}$ (electronic bandgap) with change in the interlayer distance, $d$, for the XMoY bilayers. Layer-contribution-projected band structure of AB-stacked (c,d) TeMoS, (e,f) TeMoSe, and (g,h) MoTe$_2$ bilayers, with interlayer distances reduced from their respective equilibrium values by the indicated percentages. The blue and red circles represent the region around the points in the valence band and the conduction band, respectively,  which crosses the Fermi level. The blue and red colors of the bands represent the amount of contribution of the bottom and top layers to the particular band, respectively. Here, Q is the midpoint between K and $\Gamma$. The zero of the energy scale is at the Fermi level.}       
\label{fs13}
%\end{center}
\end{figure}

\begin{figure}[!htb]
%\begin{center}
\includegraphics[width=0.7\linewidth]{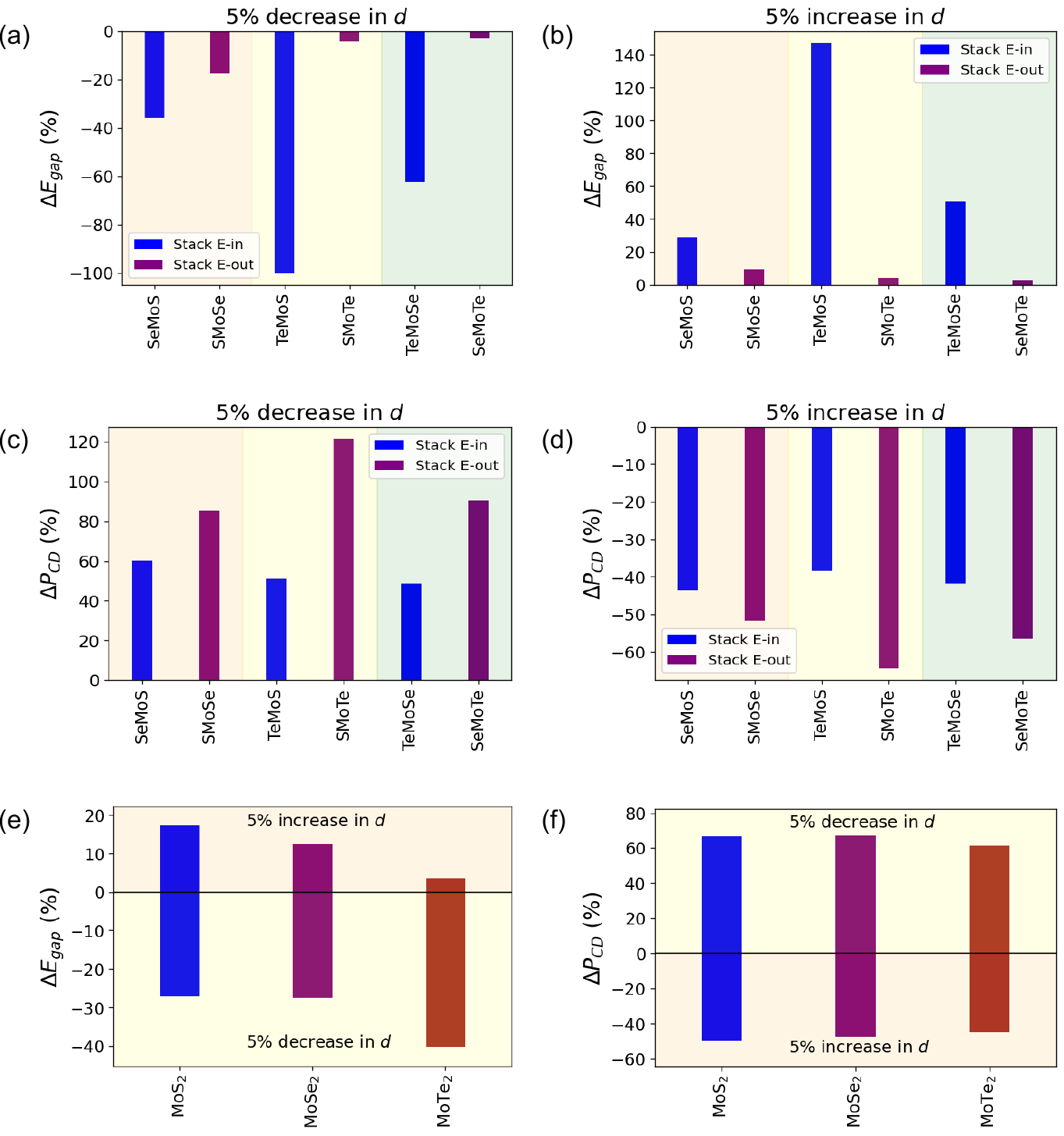}
\caption{Percentage change in (a, b) $E_{gap}$ (electronic bandgap) and (c, d) $P_{CD}$ (polarization calculated from charge density) on (a, c) 5\% decrease, and (b, d) 5\% increase in the interlayer distance, \textit{d}, for Janus XMoY bilayers. Percentage change in (e) $E_{gap}$ and (f) $P_{CD}$ on changing \textit{d} by 5\% for MoX$_2$ parent bilayers.}       
\label{fs14}
%\end{center}
\end{figure}

\begin{figure}[!htb]
%\begin{center}
\includegraphics[width=0.9\linewidth]{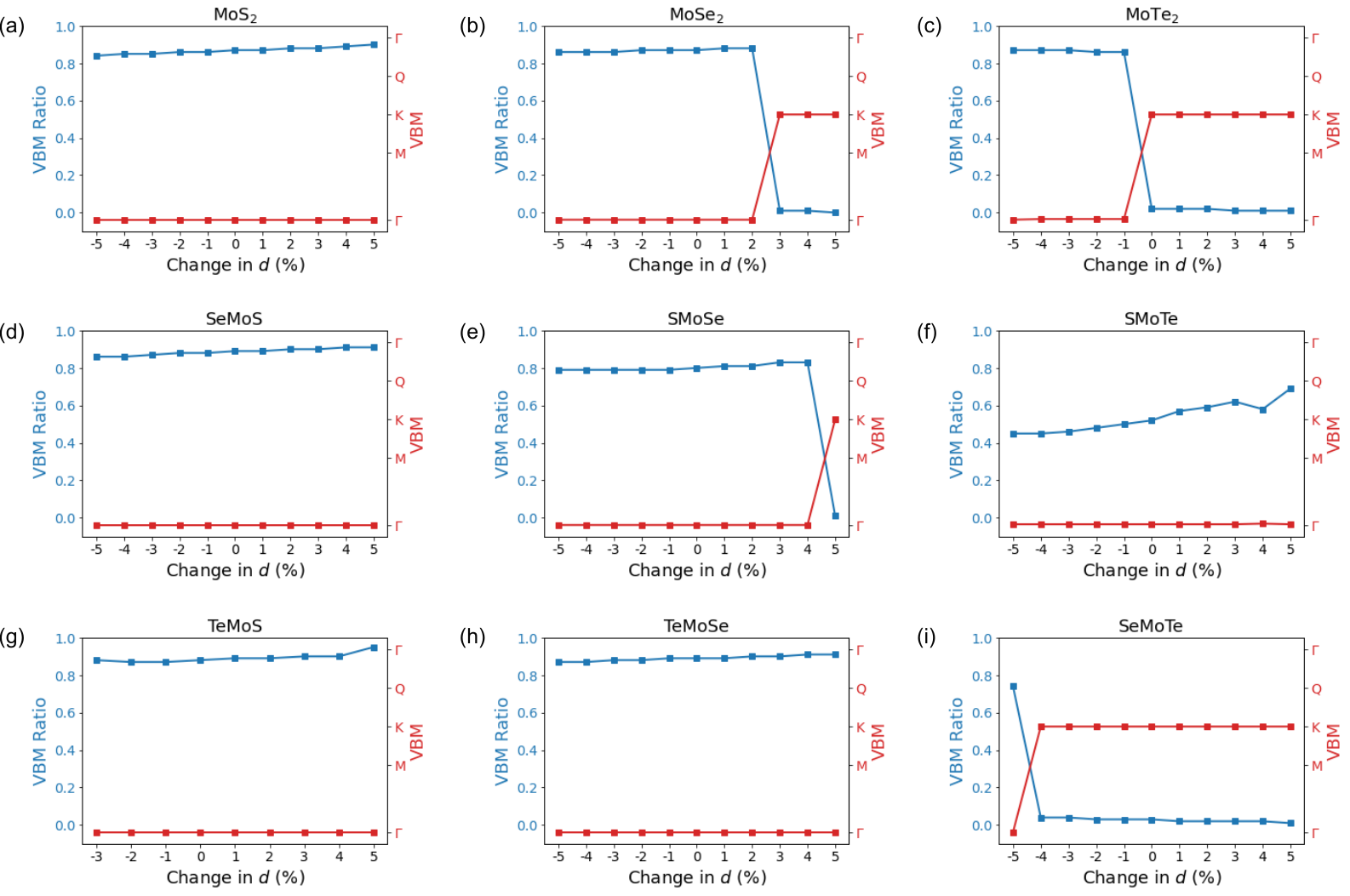}
\caption{Variation of VBM Ratio (and VBM) with change in the interlayer distance, \textit{d}, for the AB-stacked bilayers (a) MoS$_2$, (b) MoSe$_2$, (c) MoTe$_2$, (d) SeMoS, (e) SMoSe, (f) SMoTe, (g) TeMoS, (h) TeMoSe, and (i) SeMoTe. Here, VBM stands for valence band maximum.}       
\label{fs15}
%\end{center}
\end{figure}

\begin{figure}[!htb]
%\begin{center}
\includegraphics[width=0.9\linewidth]{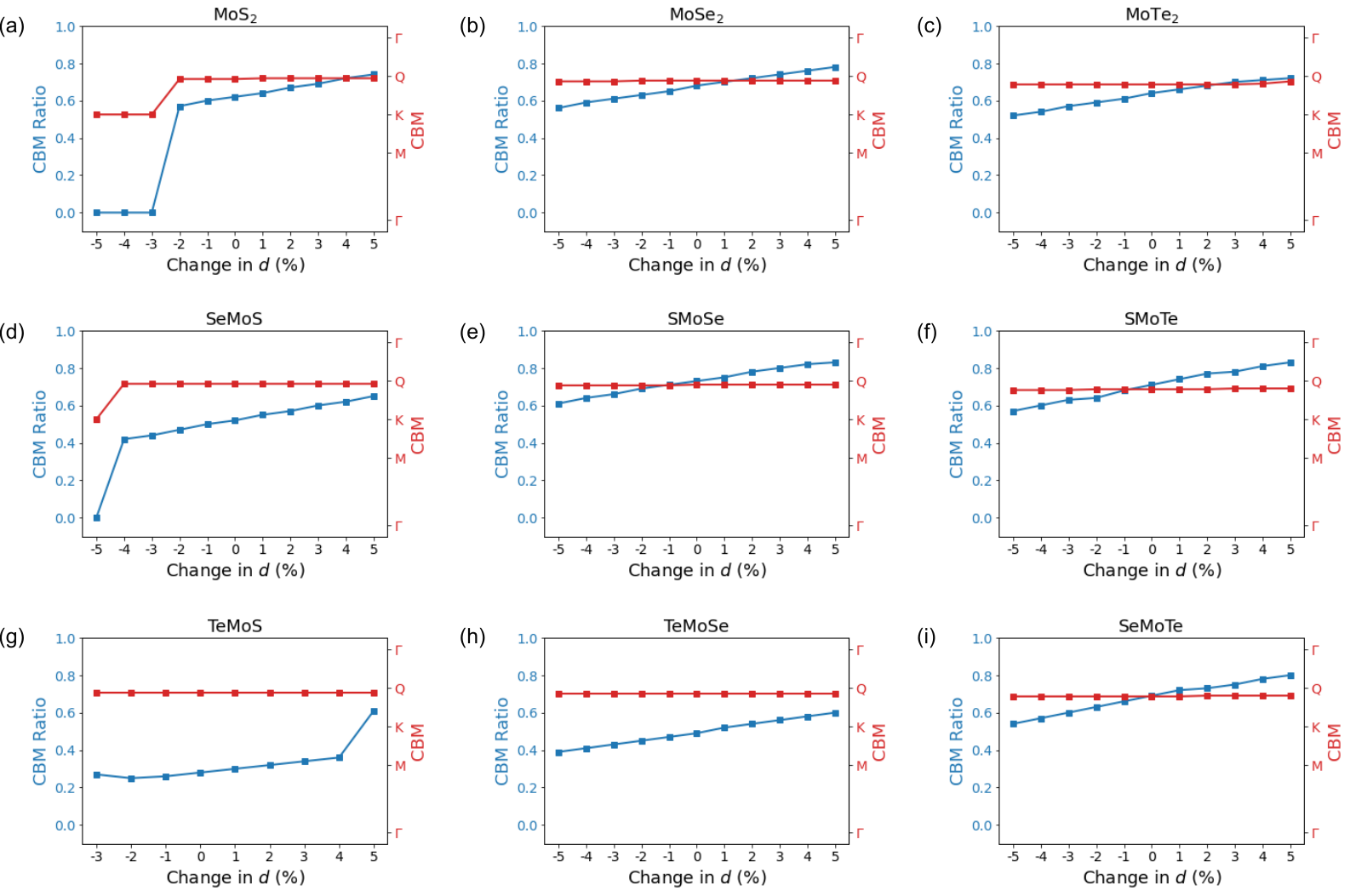}
\caption{Variation of CBM Ratio (and CBM) with change in the interlayer distance, \textit{d}, for the AB-stacked bilayers (a) MoS$_2$, (b) MoSe$_2$, (c) MoTe$_2$, (d) SeMoS, (e) SMoSe, (f) SMoTe, (g) TeMoS, (h) TeMoSe, and (i) SeMoTe. Here, CBM stands for conduction band minimum.}       
\label{fs16}
%\end{center}
\end{figure}

\begin{figure}[!htb]
%\begin{center}
\includegraphics[width=0.7\linewidth]{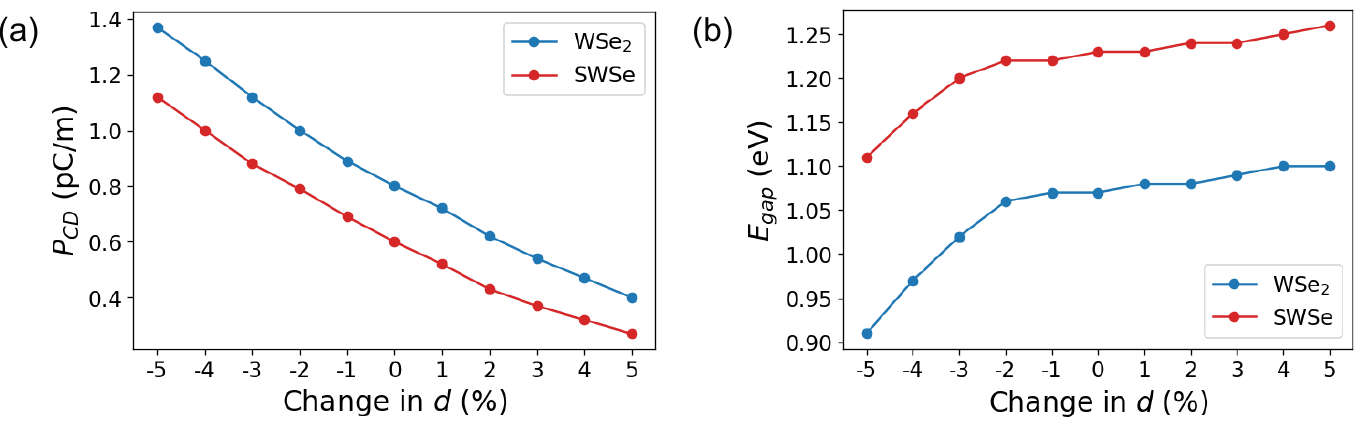}
\caption{Variation of (a) $P_{CD}$ (polarization calculated from charge density), and (b) $E_{gap}$ (electronic bandgap) with the change in the interlayer distance, \textit{d}, for WSe$_2$ and SWSe bilayers.}       
\label{fs17}
%\end{center}
\end{figure}

\begin{figure}[!htb]
%\begin{center}
\includegraphics[width=0.7\linewidth]{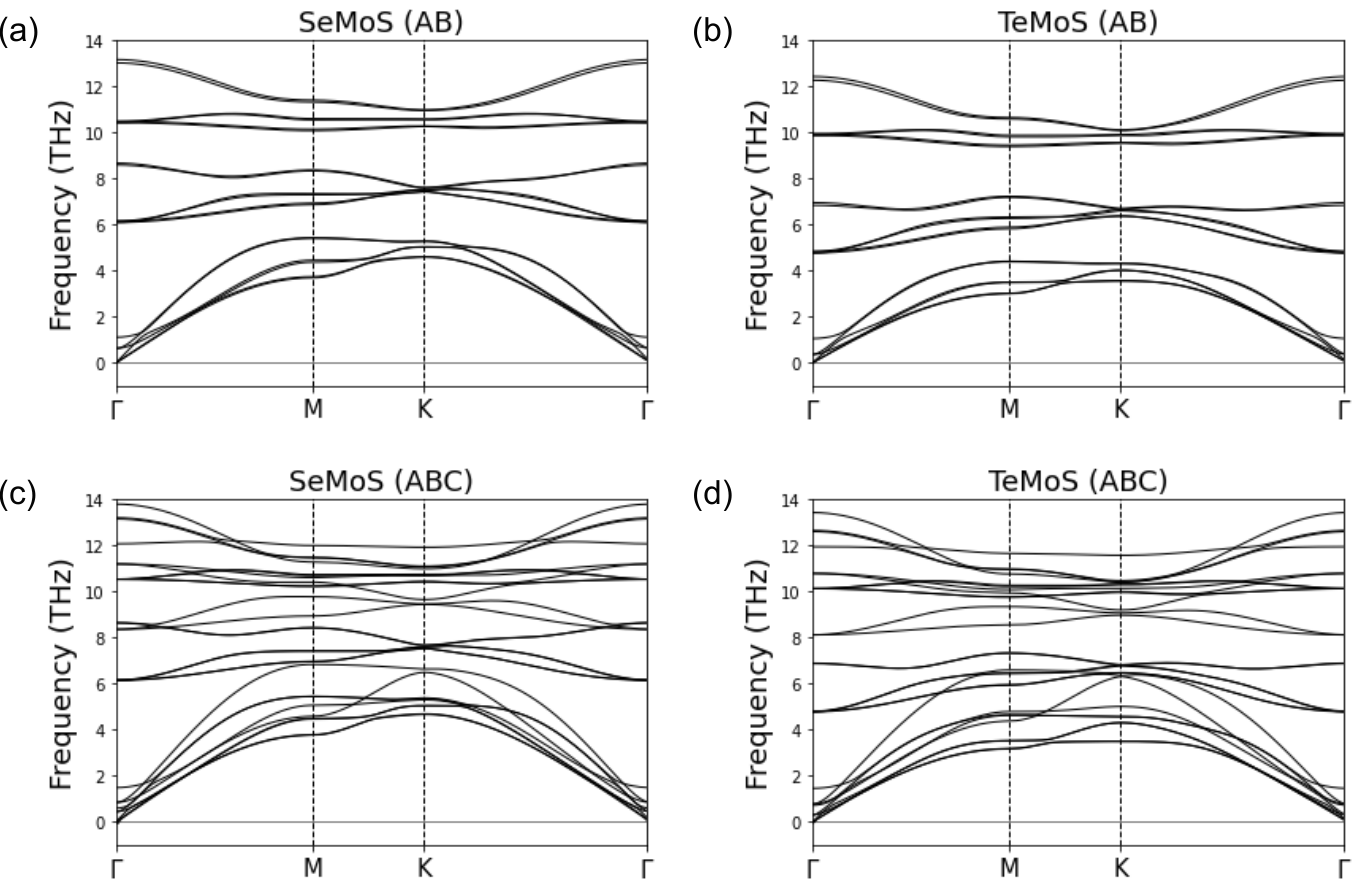}
\caption{Phonon dispersion curves for (a) SeMoS AB-stacked bilayer, (b) TeMoS AB-stacked bilayer, (c) SeMoS ABC-stacked trilayer, and (d) TeMoS ABC-stacked trilayer.}       
\label{fs18}
%\end{center}
\end{figure}

\begin{figure}[!htb]
%\begin{center}
\includegraphics[width=\linewidth]{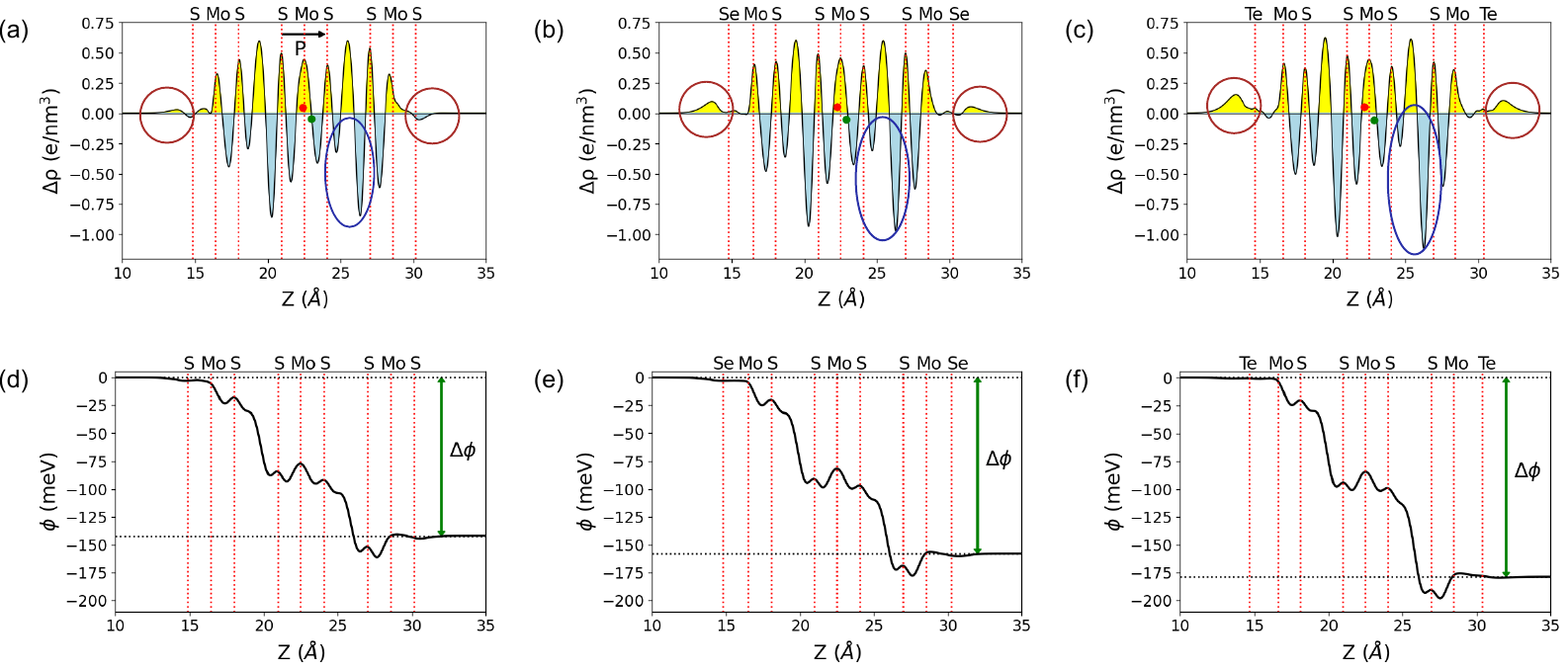}
\caption{(a-c) Charge density and (d-f) potential profiles for ABC-stacked (a, d) MoS$_2$, (b, e) SeMoS, and (c, f) TeMoS trilayer. The yellow and blue color represents the electron and hole accumulation, respectively. The polarization direction is from the red dot (pseudo electron) to the green dot (pseudo hole), as shown in (a). The red circle encloses the outer regions of the trilayers to highlight the variation in electron accumulation, and the blue circle encloses the region of variation in the hole accumulation peaks in (a-c). The $\Delta\phi$ in (d-f) represents the interlayer potential difference. The dashed red lines represent the vertical location of the atoms.}       
\label{fs19}
%\end{center}
\end{figure}

\begin{figure}[!htb]
%\begin{center}
\includegraphics[width=0.8\linewidth]{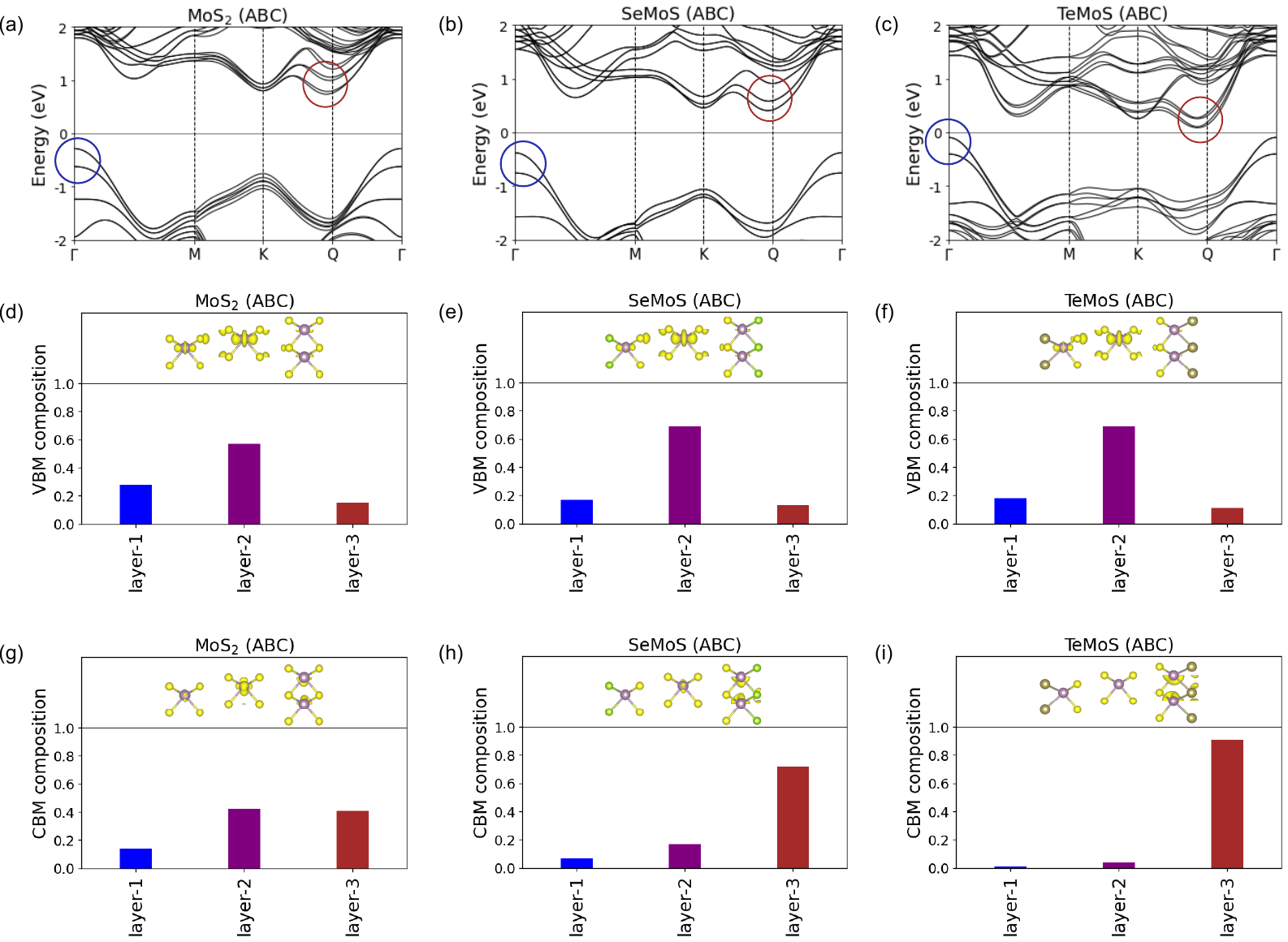}
\caption{Electronic band structure of ABC-stacked (a) MoS$_2$, (b) SeMoS, and (c) TeMoS trilayer. The blue and red circles represent the region around the VBM and the CBM, respectively. Here, Q is the midpoint between K and $\Gamma$. The zero of the energy scale is at the Fermi level. Layer-wise contribution of (d-f) VBM and (g-i) CBM for ABC-stacked (d, g) MoS$_2$, (e, h) SeMoS, and (f, i) TeMoS trilayers. The illustrations on the top in (d-i) are the partial charge density in the VBM (d-f) and CBM (g-i). Here, VBM and CBM stand for valence band maximum and conduction band minimum, respectively. }       
\label{fs20}
%\end{center}
\end{figure}

This supplementary material contains the following: (1) Formation energy plot for bilayers [Figure~\ref{fs1}] (2) charge density and potential profiles for AB and BA stacking of SeMoS bilayer [Figure~\ref{fs2}] (3) charge density and potential profiles for XMoY bilayers [Figure~\ref{fs3} and ~\ref{fs4}] (4) correlation plots for $\Delta\phi$ and $P_{CD}$ with $P_{berry}$ along with comparison of SOC and without SOC results [Figure~\ref{fs5}] (5) additional results for XMoY bilayers with $d=d_{avg}$ [Figure~\ref{fs6}] (6) band structures of XMoY bilayers [Figure~\ref{fs7}] (7) comparison of band structure of MoTe$_2$, WSe$_2$ and SWSe bilayer without and with SOC [Figure~\ref{fs8}] (8) effect of doping on polarization for bilayers [Figure~\ref{fs9} and ~\ref{fs10}] (9) variation of Fermi energy and DOS with doping for TeMoS and SMoTe bilayers [Figure~\ref{fs11} and~\ref{fs12}] (10) polarization and bandgap variation with more than 5\% reduction in interlayer distance for XMoY bilayers along with band structures for bilayers showing semi-metallic nature at reduced interlayer distance [Figure~\ref{fs13}] (11) comparison of percentage change in bandgap and polarization with change in $d$ for XMoY bilayers [Figure~\ref{fs14}] (12) variation of the VBM and CBM ratios with change in $d$ for XMoY bilayers [Figure~\ref{fs15} and~\ref{fs16}] (13) variation of polarization and bandgap with change in $d$ for WSe$_2$ and SWSe bilayers [Figure~\ref{fs17}] (14) phonon dispersion curves [Figure~\ref{fs18}] (15) charge density and potential profiles for trilayers [Figure~\ref{fs19}] (16) band structure with layer-wise contribution analysis for trilayers [Figure~\ref{fs20}] (17) tables for data for bilayer and trilayer systems presented and referred in the paper [Table~\ref{ts1},~\ref{ts2},~\ref{ts3}, and~\ref{ts4}] (18) methodology for $P_{CD}$ calculation [Section~\ref{PCD_mechanism}] (19) thermal stability analysis [Section~\ref{TStab}].

\section*{Figures}

%%%% Figure S1 %%%%
%%%% Figure S1 %%%%

%%%% Figure S2 %%%%
%%%% Figure S2 %%%%

%%%% Figure S3 %%%%
%%%% Figure S3 %%%%

%%%% Figure S4 %%%%
%%%% Figure S4 %%%%

%%%% Figure S5 %%%%
%%%% Figure S5 %%%%

%%%% Figure S6 %%%%
%%%% Figure S6 %%%%

%%%% Figure S7 %%%%
%%%% Figure S7 %%%%

%%%% Figure S8 %%%%
%%%% Figure S8 %%%%

%%%% Figure S9 %%%%
%%%% Figure S9 %%%%

%%%% Figure S10 %%%%
%%%% Figure S10 %%%%

%%%% Figure S11 %%%%
%%%% Figure S11 %%%%

%%%% Figure S12 %%%%
%%%% Figure S12 %%%%

%%%% Figure S13 %%%%
%%%% Figure S13 %%%%

%%%% Figure S14 %%%%
%%%% Figure S14 %%%%

%%%% Figure S15 %%%%
%%%% Figure S15 %%%%

%%%% Figure S16 %%%%
%%%% Figure S16 %%%%

%%%% Figure S17 %%%%
%%%% Figure S17 %%%%

%%%% Figure S18 %%%%
%%%% Figure S18 %%%%

%%%% Figure S19 %%%%
%%%% Figure S19 %%%%

%%%% Figure S20 %%%%
%%%% Figure S20 %%%%

\clearpage

\section*{Tables}

%%%% Table S1 %%%% 
\begin{table}[H]
\centering
\resizebox{\linewidth}{!}{
\begin{tabular}{||l | c | wc{2cm} | wc{2cm} | c | c | c||} 
 \hline
 \multicolumn{1}{||l|}{\textbf{Material}} & \multicolumn{1}{c|}{\textbf{Lattice Constant (\AA)}} & \multicolumn{2}{c|}{\textbf{Bond Length (\AA)}} & \multicolumn{1}{c|}{\textbf{Interlayer Distance (\AA)}} & \multicolumn{1}{c|}{\textbf{Interface Distance (\AA)}} & \multicolumn{1}{c||}{\textbf{Formation Energy (eV/atom)}} \\ [0.5ex]
 %\textbf{Material} & \textbf{Lattice Parameter (\AA)} & \textbf{Bond Length (\AA)} &  & \textbf{Interlayer Distance (\AA)} & \textbf{Interface Distance (\AA)} & \textbf{Formation Energy (eV/f.u.)} \\ [0.5ex]
 \hline
  \textbf{XMY} & \textit{\textbf{a }}$\mathbf{=}$\textit{\textbf{ b}} & \textbf{M-X} & \textbf{M-Y} & \textbf{\textit{d}} & \textbf{\textit{t}} & \textbf{$\mathbf{E_F}$} \\ [0.5ex]
 \hline\hline
 %% XMoY systems %%
 \textcolor{red}{\textbf{MoS$\mathbf{_2}$}}   & 3.158 & 2.404 & 2.404 & 6.086 & 2.956 & -1.27 \\
 \hline
 \textcolor{red}{\textbf{SeMoS}}              & 3.221 & 2.526 & 2.409 & 5.971 & 2.909 & -1.17 \\
 \hline
 \textcolor{red}{\textbf{TeMoS}}              & 3.335 & 2.714 & 2.425 & 5.802 & 2.854 & -0.92 \\
 \hline
 \textcolor{red}{\textbf{MoSe$\mathbf{_2}$}}  & 3.286 & 2.530 & 2.530 & 6.415 & 3.067 & -1.08 \\
 \hline
 \textcolor{red}{\textbf{SMoSe}}              & 3.221 & 2.410 & 2.526 & 6.538 & 3.122 & -1.17 \\
 \hline
 \textcolor{red}{\textbf{TeMoSe}}             & 3.400 & 2.714 & 2.544 & 6.214 & 2.978 & -0.87 \\
 \hline
 \textcolor{red}{\textbf{MoTe$\mathbf{_2}$}}  & 3.511 & 2.722 & 2.722 & 6.939 & 3.302 & -0.71 \\
 \hline
 \textcolor{red}{\textbf{SMoTe}}              & 3.332 & 2.425 & 2.716 & 7.284 & 3.454 & -0.93 \\
 \hline
 \textcolor{red}{\textbf{SeMoTe}}             & 3.397 & 2.544 & 2.716 & 7.152 & 3.397 & -0.87 \\
 \hline
 %% XWY systems %%
 \textcolor{blue}{\textbf{WS$\mathbf{_2}$}}   & 3.163 & 2.412 & 2.412 & 6.116 & 2.969 & -1.19 \\
 \hline
 \textcolor{blue}{\textbf{SeWS}}              & 3.225 & 2.534 & 2.417 & 6.015 & 2.935 & -1.05 \\
 \hline
 \textcolor{blue}{\textbf{TeWS}}              & 3.335 & 2.720 & 2.431 & 5.849 & 2.882 & -0.76 \\
 \hline
 \textcolor{blue}{\textbf{WSe$\mathbf{_2}$}}  & 3.290 & 2.539 & 2.539 & 6.455 & 3.088 & -0.94 \\
 \hline
 \textcolor{blue}{\textbf{SWSe}}              & 3.225 & 2.418 & 2.533 & 6.570 & 3.136 & -1.06 \\
 \hline
 \textcolor{blue}{\textbf{TeWSe}}             & 3.402 & 2.722 & 2.552 & 6.275 & 3.018 & -0.68 \\
 \hline
 \textcolor{blue}{\textbf{WTe$\mathbf{_2}$}}  & 3.515 & 2.730 & 2.730 & 6.974 & 3.322 & -0.49 \\
 \hline
 \textcolor{blue}{\textbf{SWTe}}              & 3.333 & 2.432 & 2.722 & 7.295 & 3.447 & -0.77 \\
 \hline
 \textcolor{blue}{\textbf{SeWTe}}             & 3.400 & 2.552 & 2.723 & 7.176 & 3.405 & -0.68 \\
 \hline
\end{tabular}}
\caption{Lattice constant ($a=b$), bond lengths (M-X and M-Y), interlayer distance (\textit{d}), interface distance ($t$), and formation energy per atom in the unit cell ($E_F$) for AB-stacked TMD bilayers (MX$_2$) and their Janus counterparts (XMY). Note that all the structures have polar spacegroup \textit{P3m1}.} 
\label{ts1}
\end{table}
%%%% Table S1 %%%%

%%%% Table S2 %%%% 
\begin{table}[H]
\centering
\resizebox{0.7\linewidth}{!}{
\begin{tabular}{||l | c | c | c | c | wc{1.5cm} | wc{1.5cm}||} 
 \hline
 \multicolumn{1}{||l|}{\textbf{Material}} & \multicolumn{1}{c|}{\textbf{P$\mathbf{_{berry}}$ (pC/m)}} &  \multicolumn{1}{c|}{\textbf{P$\mathbf{_{CD}}$ (pC/m)}} & \multicolumn{1}{c|}{\textbf{$\mathbf{\Delta\phi}$ (meV)}} & \multicolumn{1}{c|}{\textbf{E$\mathbf{_{gap}}$(eV)}} & \multicolumn{2}{c||}{\textbf{Depolarization (\%)}} \\ [0.5ex]
 \hline
  \textbf{XMY} &  &  &  &  & \textbf{hole} & \textbf{electron} \\ [0.5ex]
 \hline\hline
 %% XMoY systems %%
 \textcolor{red}{\textbf{MoS$\mathbf{_2}$}}   & 0.88 (0.70) & 0.84 (0.62) & 70.34 & 1.22 & 3.85 & 27.54 \\
 \hline
 \textcolor{red}{\textbf{SeMoS}}              & 1.04 (0.73) & 1.04 (0.76)  & 85.86 & 0.86 & 3.41 & 31.53 \\
 \hline
 \textcolor{red}{\textbf{TeMoS}}              & 1.25 (0.91) & 1.26 (0.89) & 100.71 & 0.17 & 1.85 & 36.15 \\
 \hline
 \textcolor{red}{\textbf{MoSe$\mathbf{_2}$}}  & 0.73 (0.52) & 0.82 (0.59) & 67.21 & 1.13 & 1.82 & 18.43 \\
 \hline
 \textcolor{red}{\textbf{SMoSe}}              & 0.61 (0.44) & 0.60 (0.44)  & 49.49 & 1.26 & 15.84 & 23.43 \\
 \hline
 \textcolor{red}{\textbf{TeMoSe}}             & 1.13 (0.74) & 1.15 (0.80) & 90.81 & 0.53 & 0.82 & 22.09 \\
 \hline
 \textcolor{red}{\textbf{MoTe$\mathbf{_2}$}}  & 0.93 (0.74) & 1.00 (0.71)  & 79.83 & 0.87 & 19.22 & 12.05 \\
 \hline
 \textcolor{red}{\textbf{SMoTe}}              & 0.42 (0.18) & 0.42 (0.31) & 35.61 & 0.89 & 42.79 & 22.19 \\
 \hline
 \textcolor{red}{\textbf{SeMoTe}}             & 0.54 (0.31) & 0.62 (0.46) & 51.92 & 1.02 & 79.18 & 14.79 \\
 \hline
 %% XWY systems %%
 \textcolor{blue}{\textbf{WS$\mathbf{_2}$}}   & 0.82 (0.64) & 0.85 (0.62) & 69.83 & 1.26 & 7.57 & 31.29 \\
 \hline
 \textcolor{blue}{\textbf{SeWS}}              & 1.02 (0.78) & 1.00 (0.72) & 81.86 & 1.08 & 4.75 & 31.99 \\
 \hline
 \textcolor{blue}{\textbf{TeWS}}              & 1.15 (0.73) & 1.20 (0.85) & 95.90 & 0.29 & 1.38 & 33.48 \\
 \hline
 \textcolor{blue}{\textbf{WSe$\mathbf{_2}$}}  & 0.75 (0.60) & 0.80 (0.56) & 63.01 & 1.07 & 64.39 & 26.88 \\
 \hline
 \textcolor{blue}{\textbf{SWSe}}              & 0.58 (0.16) & 0.60 (0.43) & 48.78 & 1.23 & 54.71 & 29.96 \\
 \hline
 \textcolor{blue}{\textbf{TeWSe}}             & 1.01 (0.78) & 1.08 (0.76) & 85.90 & 0.74 & 1.86 & 25.08 \\
 \hline
 \textcolor{blue}{\textbf{WTe$\mathbf{_2}$}}  & 0.94 (0.75) & 0.99 (0.68) & 77.30 & 0.79 & 59.33 & 15.70 \\
 \hline
 \textcolor{blue}{\textbf{SWTe}}              & 0.35 (0.13) & 0.39 (0.26) & 29.25 & 0.96 & 32.17 & 32.20 \\
 \hline
 \textcolor{blue}{\textbf{SeWTe}}             & 0.56 (0.37) & 0.58 (0.42) & 47.26 & 0.91 & 115.61 & 21.68 \\
 \hline
\end{tabular}}
\caption{Polarization via berry phase method ($P_{berry}$), polarization from charge density ($P_{CD}$), interlayer potential difference ($\Delta\phi$), electronic bandgap($E_{gap}$), and depolarization via hole and electron doping for AB-stacked TMD bilayers (MX$_2$) and their Janus counterparts (XMY). The values in parentheses for $P_{berry}$ and $P_{CD}$ include dipole corrections.}
\label{ts2}
\end{table}
%%%% Table S2 %%%%

%%%% Table S3 %%%% 
\begin{table}[H]
\centering
\resizebox{\linewidth}{!}{
\begin{tabular}{||l | c | c | c | c | c | c | c | c ||} 
 \hline
 \textbf{Material} & \textbf{Lattice Constant (\AA)} & \textbf{Interface Distance (\AA)} & \textbf{P$\mathbf{_{CD}}$ (pC/m)} & \textbf{$\mathbf{l_h}$ (e/nm$^3$)} & \textbf{ $\mathbf{l_e}$ (e/nm$^3$)} & \textbf{E$\mathbf{_{gap}}$ (eV)} & \textbf{VBM Ratio} & \textbf{CBM Ratio} \\ [0.5ex]
 \hline
  \textbf{XMY} & \textit{\textbf{a }}$\bf{=}$\textit{\textbf{ b}} & $\mathbf{t}$ &  &  &  &  & & \\ [0.5ex]
 \hline\hline
 %% Without Relaxation %%
 \multicolumn{9}{||c||}{\textbf{Without Relaxation}} \\ [0.5ex]
 \hline
 \hline
 \textcolor{red}{\textbf{MoS$\mathbf{_2}$}}   & 3.15 & 3.35 & 0.35 & 0.18 & 0.32 & 1.49 & 0.91 ($\Gamma$) & 0.78 (Q)  \\
 \hline
 \textcolor{red}{\textbf{SeMoS}}              & 3.15 & 3.35 & 0.33 & 0.15 & 0.28 & 1.56 & 0.92 ($\Gamma$) & 0.79 (Q$'$) \\
 \hline
 \textcolor{red}{\textbf{TeMoS}}              & 3.15 & 3.35 & 0.30 & 0.16 & 0.31 & 1.60 & 0 (K) & 0.78 (Q$'$) \\
 \hline
 \textcolor{red}{\textbf{MoSe$\mathbf{_2}$}}  & 3.15 & 3.35 & 0.57 & 0.29 & 0.69 & 1.46 & 0.91 ($\Gamma$) & 0 (K) \\
 \hline
 \textcolor{red}{\textbf{SMoSe}}              & 3.15 & 3.35 & 0.59 & 0.31 & 0.73 & 1.39 & 0.90 ($\Gamma$) & 0.71 (Q$'$) \\
 \hline
 \textcolor{red}{\textbf{TeMoSe}}             & 3.15 & 3.35 & 0.54 & 0.27 & 0.76 & 1.51 & 0.93 ($\Gamma$) & 0.79  (Q$'$) \\
 \hline
 \textcolor{red}{\textbf{MoTe$\mathbf{_2}$}}  & 3.15 & 3.35 & 1.34 & 0.52 & 1.60 & 0.78 & 0.88 (P) & 0 (K) \\
 \hline
 \textcolor{red}{\textbf{SMoTe}}              & 3.15 & 3.35 & 1.35 & 0.62 & 1.55 & 1.03 & 0.86 (P) & 0 (K) \\
 \hline
 \textcolor{red}{\textbf{SeMoTe}}             & 3.15 & 3.35 & 1.36 & 0.59 & 1.50 & 0.98 & 0.88 (P) & 0 (K) \\
 \hline
 %% Constrained Relaxation %%
 \multicolumn{9}{||c||}{\textbf{Constrained Relaxation}} \\ [0.5ex]
 \hline
 \hline
 \textcolor{blue}{\textbf{MoS$\mathbf{_2}$}}   & 3.15 & 3.35 & 0.35 & 0.18 & 0.32 & 1.49 & 0.91 ($\Gamma$) & 0.78 (Q) \\
 \hline
 \textcolor{blue}{\textbf{SeMoS}}              & 3.22 & 3.42 & 0.34 & 0.17 & 0.22 & 1.25 & 0.93 ($\Gamma$) & 0.73 (Q$'$) \\
 \hline
 \textcolor{blue}{\textbf{TeMoS}}              & 3.33 & 3.52 & 0.31 & 0.13 & 0.11 & 0.64 & 0.94 ($\Gamma$) & 0.57 (Q$'$) \\
 \hline
 \textcolor{blue}{\textbf{MoSe$\mathbf{_2}$}}  & 3.29 & 3.14 & 0.73 & 0.40 & 0.95 & 1.18 & 0.88 ($\Gamma$) & 0.7 (Q$'$) \\
 \hline
 \textcolor{blue}{\textbf{SMoSe}}              & 3.22 & 3.08 & 0.67 & 0.40 & 1.07 & 1.23 & 0.79 ($\Gamma$) & 0.72 (Q$'$) \\
 \hline
 \textcolor{blue}{\textbf{TeMoSe}}             & 3.40 & 3.24 & 0.72 & 0.37 & 0.79 & 0.77 & 0.91 ($\Gamma$) & 0.58 (Q$'$) \\
 \hline
 \textcolor{blue}{\textbf{MoTe$\mathbf{_2}$}}  & 3.53 & 2.92 & 1.81 & 1.01 & 2.28 & 0.36 & 0.87 ($\Gamma$) & 0.01 (K) \\
 \hline
 \textcolor{blue}{\textbf{SMoTe}}              & 3.37 & 2.80 & 1.70 & 0.68 & 2.59 & 0.65 & 0.62 ($\Gamma$) & 0.22 (K) \\
 \hline
 \textcolor{blue}{\textbf{SeMoTe}}             & 3.43 & 2.84 & 1.74 & 0.85 & 2.44 & 0.65 & 0.81 ($\Gamma$) & 0.01 (K) \\
 \hline
\end{tabular}}
\caption{Results for the experiment with a fixed interlayer distance ($d=d_{avg}=6.49$\AA) under both without relaxation and constrained relaxation conditions for XMoY bilayers: lattice parameter ($a=b$), interface distance($t$), polarization from charge density ($P_{CD}$), charge density profile parameters ($l_h$ and $l_e$), electronic bandgap ($E_{gap}$), and VBM (valence band maximum) and CBM (conduction band minimum) ratios. Here, the positions of the VBM and CBM are provided in the brackets in the VBM and CBM Ratios columns. Q$'$ represents some k point near the high-symmetry point Q, which is the midpoint between K and $\Gamma$.}
\label{ts3}
\end{table}
%%%% Table S3 %%%%

%%%% Table S4 %%%% 
\begin{table}[H]
\centering
\resizebox{\linewidth}{!}{
\begin{tabular}{||l | c | wc{2cm} | wc{2cm} | c | c | c | c | c||} 
 \hline
 \multicolumn{1}{||l|}{\textbf{Material}} & \multicolumn{1}{c|}{\textbf{Lattice Constant (\AA)}} & \multicolumn{2}{c|}{\textbf{Interlayer Distance (\AA)}} & \multicolumn{1}{c|}{\textbf{$\mathbf{P_{berry}}$ (pC/m)}} & \multicolumn{1}{c|}{\textbf{$\mathbf{P_{CD}}$ (pC/m)}} & \multicolumn{1}{c|}{\textbf{$\mathbf{\Delta\phi}$ (meV)}} & \multicolumn{1}{c|}{\textbf{$\mathbf{E_{gap}}$ (eV)}} &  \multicolumn{1}{c||}{\textbf{E$\mathbf{_F}$(eV/atom)}} \\ [0.5ex]
 \hline
  \textbf{XMY} & \textit{\textbf{a }}$\bf{=}$\textit{\textbf{ b}} & \textbf{d$\mathbf{_1}$} & \textbf{d$\mathbf{_2}$} &  &  & & & \\ [0.5ex]
 \hline\hline
 %% XMoY systems %%
 \textcolor{red}{\textbf{MoS$\mathbf{_2}$}}   & 3.16 & 6.09 & 6.09 & 1.84 (1.15) & 1.91 (1.26) & 142.20 & 1.02 & -1.29 \\
 \hline
 \textcolor{red}{\textbf{SeMoS}}              & 3.20 & 6.01 & 6.02 & 2.15 (1.41) & 2.16 (1.40) & 157.82 & 0.79 & -1.21 \\
 \hline
 \textcolor{red}{\textbf{TeMoS}}              & 3.27 & 5.91 & 5.91 & 2.48 (1.71) & 2.54 (1.58) & 178.77 & 0.18 & -1.03 \\
 \hline
\end{tabular}}
\caption{Lattice constant ($a=b$), interlayer distances ($d_1$ and $d_2$), polarization via berry phase method ($P_{berry}$), polarization from charge density ($P_{CD}$), interlayer potential difference ($\Delta\phi$), electronic bandgap($E_{gap}$) and formation energy per atom in the unit cell ($E_F$) for ABC-stacked XMoS (X=S,Se,Te) trilayers. Note that all the structures have polar spacegroup \textit{P3m1}. The values in parentheses for $P_{berry}$ and $P_{CD}$ include dipole corrections.}
\label{ts4}
\end{table}
%%%% Table S4 %%%%

%%%% Polarization Calculation using Charge Density Plots %%%%
\section{Polarization calculation from charge density ($P_{CD}$)}
\label{PCD_mechanism}

The polarization values are calculated using the charge density profiles of the bilayers. To obtain the polarization from the charge density profiles, we follow the steps outlined below.
\begin{itemize}
    \item Obtain the pseudo electron ($q_e$) and hole ($q_h$) charges by taking the average of the yellow and blue curves, respectively, and multiplying them by the volume of the unit cell. Here, these pseudo charges, $q_e$ and $q_h$, represent the negative and positive charges of the dipole, respectively.  
    \item Then, we determine the center of mass of the yellow and blue curves, representing the position of the pseudo electron ($z_e$) and pseudo hole ($z_h$) charges along the $z$ direction. The center of mass ($z_{COM}$) of a charge density curve ($\Delta\rho(z)$) is obtained by using the formula:
    \begin{equation}
        z_{COM} = \frac{\int z \Delta\rho(z) \,dz}{\int \Delta\rho(z) \,dz}.
    \label{eqzCOM}
    \end{equation}
    \item The polarization is calculated by making an electric dipole moment using the pseudo charges obtained from the charge density profiles and calculating the dipole moment per unit area, where we take the in-plane area of the bilayers. The polarization is calculated using the formula
    \begin{equation}
       P_{CD} = \frac{\sum q_{pseudo} z_{COM}}{Area} = \frac{q_{e}z_{e}+q_{h}z_{h}}{Area}. 
    \label{eqPCD}
    \end{equation}  
\end{itemize}
The charge density profiles for the AB and BA stacking of the SeMoS bilayer are shown below, which represent how the relative position of the pseudo electron (red dot) and pseudo hole (green dot) determines the direction of the polarization. Note that the pseudo charges are quite close to the peaks of the electron (red circle) and hole (green circle) accumulation within the interlayer space. This polarization calculation methodology can be extended to trilayer systems as well, as done in this work.  
%%%% Figure P_CD %%%%
\begin{figure}[!htb]
\includegraphics[width=0.8\linewidth]{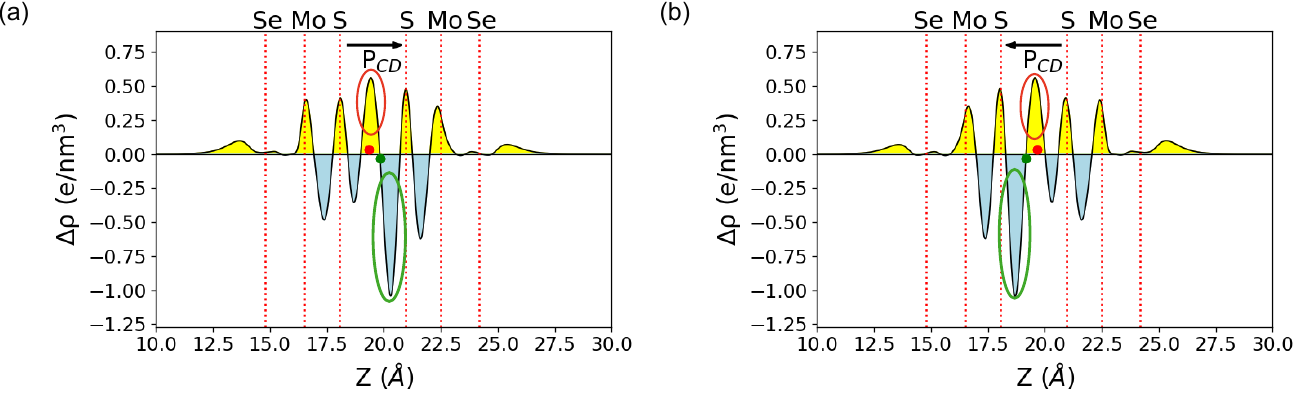}
\begin{center}
    Charge density profiles for SeMoS (a) AB and (b) BA stacking bilayer. 
\end{center}      
\label{PCD}
\end{figure}
%%%% Figure P_CD %%%%

%%%% Thermal Stability Analysis %%%%
\section{Thermal Stability Analysis}
\label{TStab}
Ab initio molecular dynamics (AIMD) simulations in the canonical ensemble were performed to assess the thermal stability. The simulations were conducted on a $2\times2\times1$ supercell for 5000 steps with a time step of 1 fs. Norm-conserving pseudopotentials~\cite{Hamann2013} were employed instead of ultrasoft pseudopotentials~\cite{Vanderbilt1990} due to their superior transferability and a simpler, more stable evaluation of forces and related properties, which is essential for reliable molecular dynamics. A kinetic energy cutoff of 90 Ry was used for the wavefunctions and 360 Ry for the charge density.

The AIMD simulation results for the bilayer (AB) and trilayer (ABC) Janus sliding ferroelectrics are provided below. No abrupt energy changes were observed during the energy evolution, indicating that these structures remain robustly stable over the simulation time at the specified temperatures.

%%%% Figure AIMD 300K %%%%
\begin{figure}[!htb]
%\begin{center}
\includegraphics[width=0.7\linewidth]{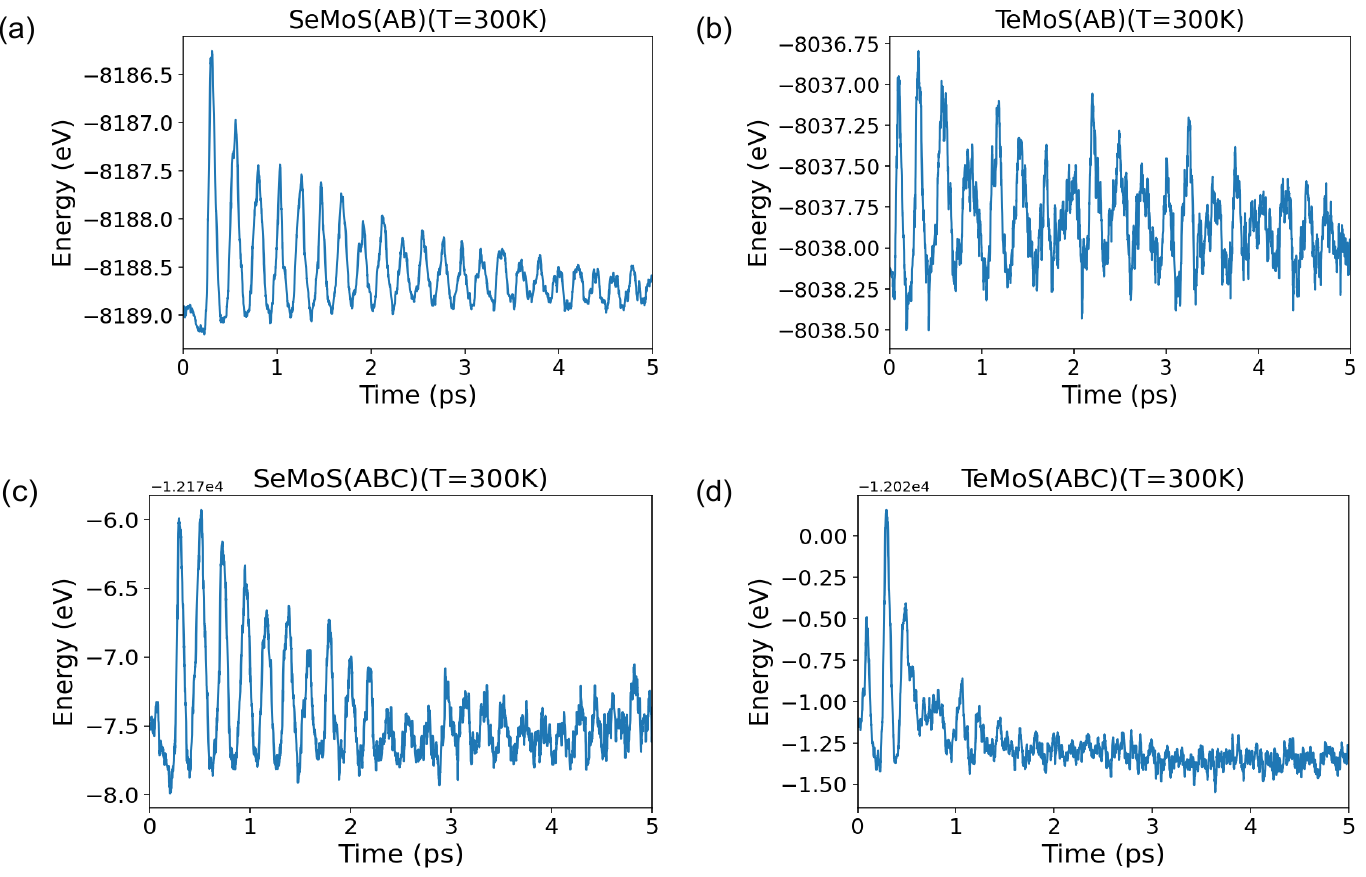}
\caption*{Ab initio MD simulations of the evolution of energy for (a) SeMoS AB-stacked bilayer, (b) TeMoS AB-stacked bilayer, (c) SeMoS ABC-stacked trilayer, and (d) TeMoS ABC-stacked trilayer, with the time running for 5 ps under 300 K.}       
\label{fAIMD300}
%\end{center}
\end{figure}
%%%% Figure AIMD 300K %%%%

%%%% Figure AIMD 600K %%%%
\begin{figure}[!htb]
%\begin{center}
\includegraphics[width=0.7\linewidth]{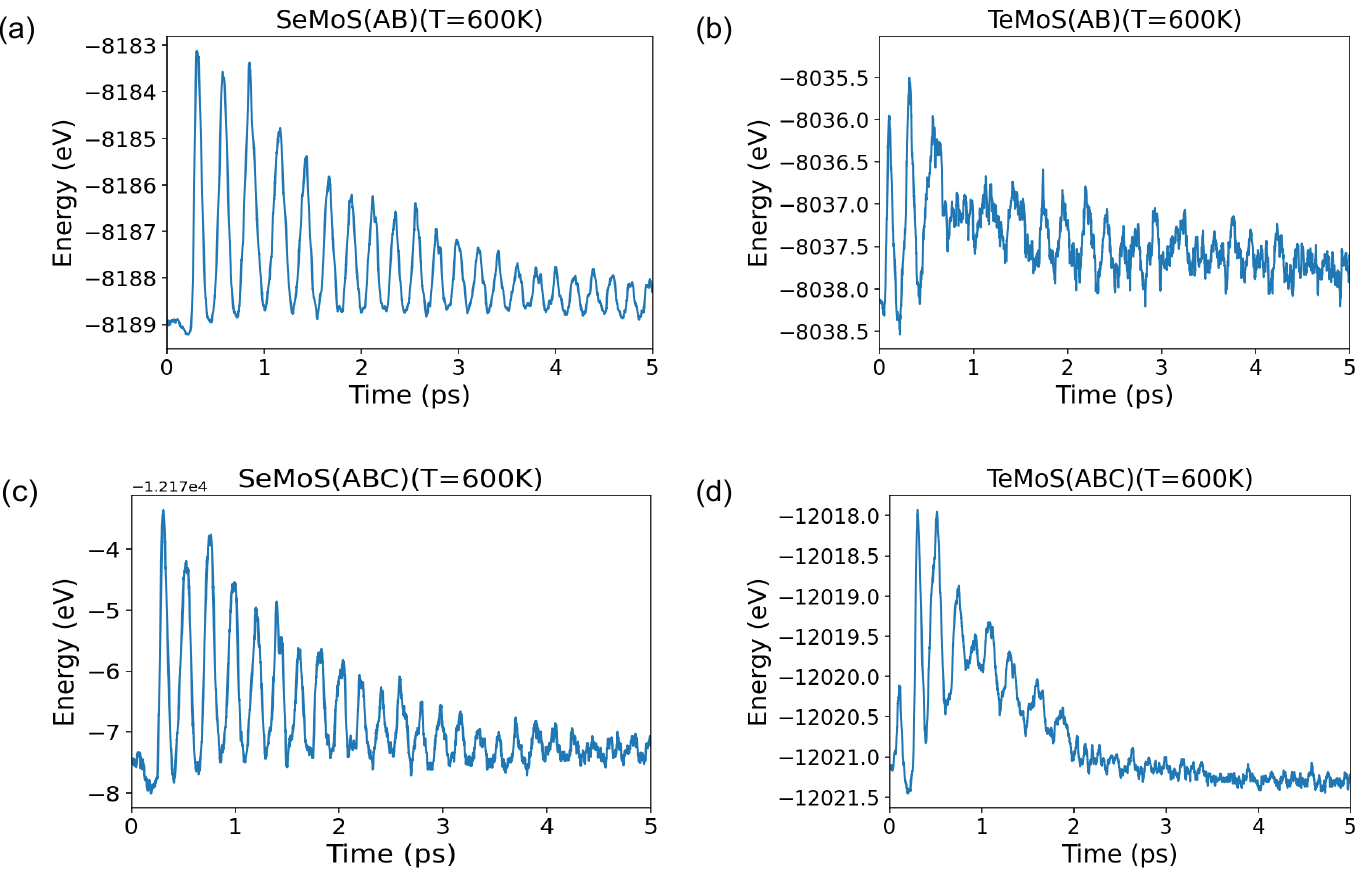}
\caption*{Ab initio MD simulations of the evolution of energy for (a) SeMoS AB-stacked bilayer, (b) TeMoS AB-stacked bilayer, (c) SeMoS ABC-stacked trilayer, and (d) TeMoS ABC-stacked trilayer, with the time running for 5 ps under 600 K.}       
\label{fAIMD600}
%\end{center}
\end{figure}
%%%% Figure AIMD 600K %%%%
\newpage

\bibliography{ref}